\documentclass[useAMS,usenatbib]{mn2e}
\usepackage{graphics,epsfig,aas_macros,amsmath}
\usepackage[]{inputenc,amssymb}

\begin{document}
\title[Variability study of IGR J18027--2016]{Variability study of the High Mass X-ray Binary IGR J18027--2016 with {\it Swift}--XRT}
\author[N. Aftab, N. Islam and B. Paul]{Nafisa Aftab $^{1}$\thanks{E-mail:nafisa@rri.res.in}, Nazma Islam$^{1,2}$ and Biswajit Paul$^{1}$ \\
$^{1}$Raman Research Institute, Sadashivnagar, Bangalore-560080, India\\
$^{2}$Joint Astronomy Programme, Indian Institute of Science, Bangalore-560012, India}

 \maketitle

 \begin{abstract}
 
We report the results from pulsations and spectral analysis of a large number of observations of the HMXB pulsar IGR J18027--2016
with {\it Swift}--XRT, carried out at different orbital phases. In some orbital phases, as seen in different XRT observations, 
the X-ray intensity is found to vary by a large factor, of about $\sim$50. In all the observations with sufficient number of
source X-ray photons, pulsations have been detected around the previously known pulse period of $\sim$140 sec,
When detected, the pulse profiles do not show any significant variation over a flux difference
of a factor of $\sim$3. The absorption column density is found to be large before and after the eclipse. We discuss
various possible reasons for intensity and spectral variations in IGR J18027--2016, such as clumpy wind and hydrodynamic instabilities.
\end{abstract}
 

\section{Introduction}
High Mass X-ray binary (HMXB) systems contain a companion star with mass $\geq$ 10 M$_{\odot}$ (either a main sequence star or a supergiant)
and a compact object (either a neutron star or a black hole). Accretion onto the compact object occur via capture of stellar wind or Roche lobe 
overflow. HMXB systems are divided into two classes:(1) Be X-ray binary (Be HMXB) and (2) Supergiant X-ray binary (sgHMXB) some of which
 show the Supergiant Fast X-ray Transient(SFXT) Phenomena. Be HMXB and SFXTs are mostly transients in nature.
In Be HMXBs, the accretion onto the compact object occurs via outflowing equatorial disk of the companion star stellar wind and the compact
object passing through it \citep{Reig2011}. Supergiant Fast X-ray Transients (SFXTs) are a sub class of HMXBs discovered with {\it INTEGRAL}, 
having recurrent, bright, short flares \citep{sguera2005}, reaching L$_{X}$ $\sim$10$^{36}-10^{37}$ erg sec$^{-1}$ \citep{romano2007},
while their quiescent X-ray luminosity is $\sim$10$^{32}$ ergs sec$^{-1}$ \citep{bozzo2010}. Persistent sgHMXBs are found to 
have X-ray luminosity L$_X = 10^{35} - 10^{36}$ ergs sec$^{-1}$, most of the time. Several short off states have been observed in
some of these systems: Vela X--1 -- \cite{mano2015, oda2013}; GX 301--2 --\cite{gog2011}; 
4U1907+09 -- \cite{doro2012}, 4U 1700--37 -- \cite{gre1999}, OAO 1657--415 -- \cite{pp2014}. On the other hand some sgHMXBs 
like SMC X-1 and LMC X-4 do not show off states, but they have strong short timescale flares \citep{moon2003_smcx-1,moon2001_lmcx4}.
INTEGRAL observation of sgHMXBs show a wide range and type of intensity variation \citep{wal2015}.

 \par
The HMXB IGR J18027--2016 was discovered with {\it INTEGRAL} -- IBIS/ISGRI during the survey 
of the Galactic Center region in September 2003 \citep{rev2004}. The pulsar is found to have a spin period of $\sim$139 sec \citep{augello2003} 
and orbital period of 4.57 days \citep{hill2005,jain2009} around a supergiant companion 
 with spectral type B1-Ib \citep{Torejon2010}. \citet{hill2005} characterised the combined {\it XMM-Newton} and {\it INTEGRAL} X-ray 
 spectrum of the pulsar by a broken power law, modified by a photoelectric absorption along the line of
 sight hydrogen column  density N$_{H}$ $\sim$10$^{23}$ cm$^{-2}$. A soft excess is also detected in the spectra of this source
 \citep{hill2005,walter2006}. 
 
 \par
In this work, we have analyzed all {\it Swift}--XRT observations of IGR J18027--2016 to investigate its long term pulsation and spectral 
characteristics. We searched for pulsations in all the observations and folded the light-curves with the estimated pulse period to study
its pulse profiles. Orbital intensity analysis show some low X-ray intensity episodes of the source, similar to that seen in Vela X--1,
GX 301--2, 4U 1907+09 \citep{mano2015,gog2011,doro2012}, OAO 1657--415 \citep{barn2008} etc. We have further investigated the nature
of the  system  by studying its spectral characteristics at different orbital phases. Our results can put some useful insight into  
systems having sudden off states in X-ray intensity.

\section{Data and Analysis}

{\it Swift} observatory was launched in November 2004 \citep{gehr2004}, consisting of 3 sets of  instruments: 1) Burst Alert Telescope (BAT), 
operating in the energy range of 15--150 keV \citep{barthelmy2005}
 2) X-ray Telescope (XRT), operating in the range of 0.2--10 keV \citep{burr2007} 3) Ultraviolet and Optical Telescope (UVOT), having UV  and 
optical coverage of 170--600 nm (Roming et al 2005). XRT and UVOT are two narrow field instruments, coaligned and pointed 
to the center of FOV of BAT. 
\par
BAT is a coded aperture instrument with  CdZnTe 
 detector, having a field of view $100^\circ$$\times$$60^\circ$ and a detection sensitivity
 of 5.3 mCrab \citep{krimm2013}. XRT is a focusing telescope which employs an X-ray CCD detector with a  Wolter 1
mirror set of 3.5 m focal length, with 23.6 $\times$ 23.6 arcmin FOV.  The imaging array consists 
of 600 $\times$ 600 image pixels, each with 40  $\mu$m$\times$  40 $\mu$m size and 2.5 arcsec/pixel resolution. 
 It operates on 3 read-out modes namely Imaging (IM),  Windowed Timing (WT) and Photon Counting (PC) mode
with few sub modes.
In Imaging mode (IM), image of the object is obtained by CCD read-out.
Photons are allowed to pile up and photon recognition is  not done in this mode. 
Windowed timing mode (WT) produces 1.7 ms resolution timing with 1D position information and full energy resolution for flux less than 
600 mCrab. 
Photon counting mode (PC) contributes to full
imaging and spectroscopic resolution with time resolution of 2.6 sec. 
\par
 We have analyzed 33 separate {\it Swift}--XRT observations of IGR J18027--2016 from MJD  54141 to 56171.
 We have also used $\sim$10.5 years of {\it Swift}--BAT lightcurve to make an accurate estimation of the 
 orbital period of this system.\footnote{ http://swift.gsfc.nasa.gov/results/transients/weak/IGRJ18027-2016}
 Minimum exposure time amongst 33 {\it Swift}--XRT observations $\sim$300 sec and   
 maximum exposure $\sim$10 kilosec.
  We used Photon Counting (PC) mode data, because 32 observations out of 33  had only this datamode.
  We filtered level1 data with the task {\small XRTPIPELINE} and obtained cleaned event files for all observations. For
 the barycenter correction of the time column of the event files we used the {\small FTOOL  BARYCORR}. We extracted the source photons
 from a region with  60$''$  radius centering the source and the background photons from a similar region in the FOV that is free of
 any other X-ray sources.  We used these  source and background region files to extract
corresponding lightcurves and spectra using {\small XSELECT} v2.4C.
\par
We have generated exposure maps with the task {\small XRTEXPOMAP}
to correct for the loss of flux due to some hot CCD pixels. We
then used this exposure  map to create ancillary file with the routine {\small\ XRTMKARF} which was then used for the spectral fitting in
{\small\ XSPEC}. We obtained the response file 
from the latest {\it Swift} calibration dataset {\small CALDB} v1.3.0.
For 5 observations, the source could not be distinctly identified from the background, even with the
task {\small\ XRTCENTROID}. For these observations, we
extracted the lightcurves and spectra with region files  centered 
 at the R.A.(18h 02m 41.94s) and Dec.($-20^\circ$ 17$'$ 17.3$''$) of the source \citep{Torejon2010}.
Observation at MJD 54141 was longer in duration ($\sim$10 ks) and observations at MJD 56085 and 56096 showed
significant difference in the count rate at the beginning and end of the observation. So we divided lightcurves and spectra of these three
observations into two parts to investigate them separately. 
Therefore, we have total 36 separate lightcurves and spectra to carry out timing and spectral analysis.

\begin{figure*}
\centering
\includegraphics[scale=0.22, angle=-90]{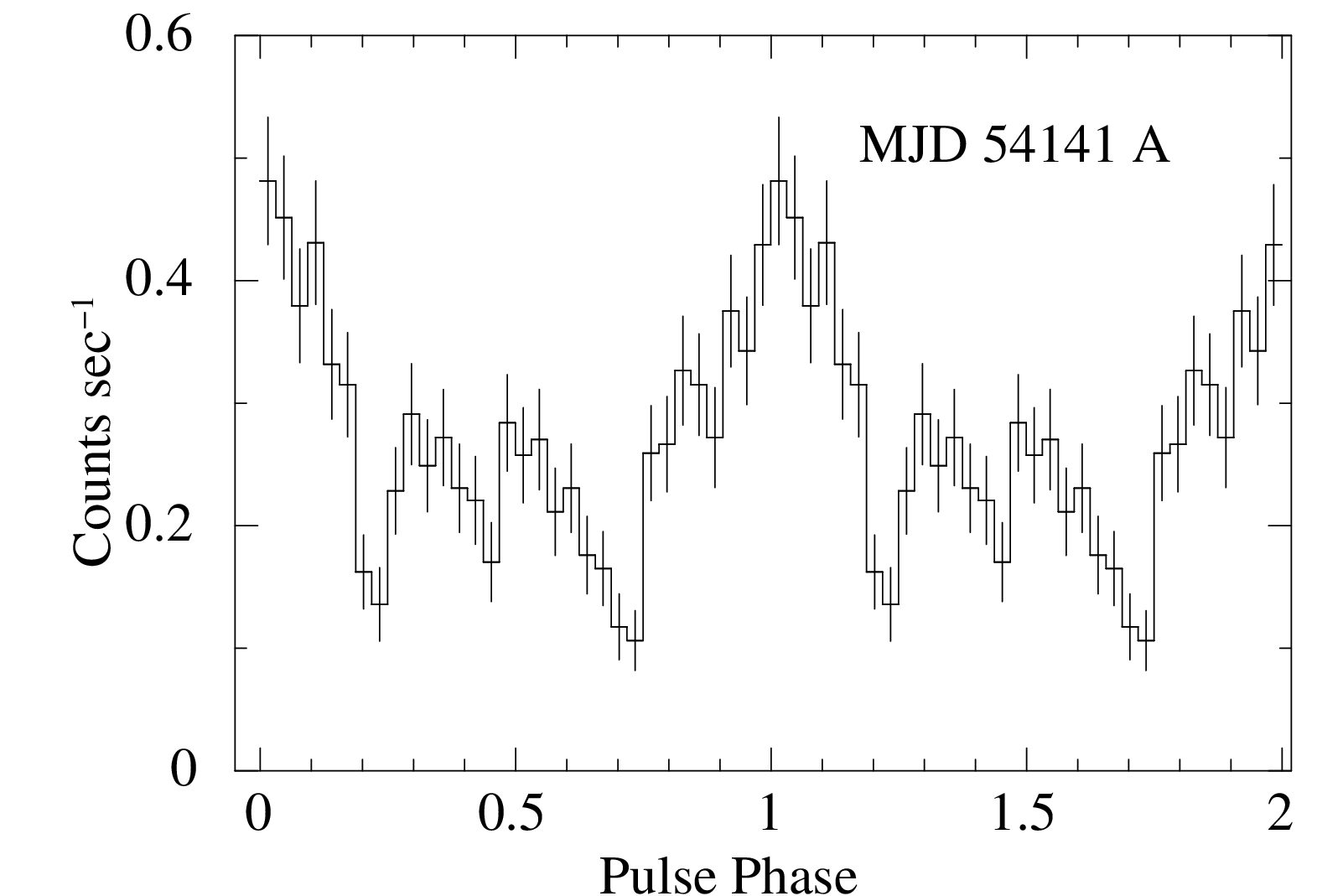}
\includegraphics[scale=0.22, angle=-90]{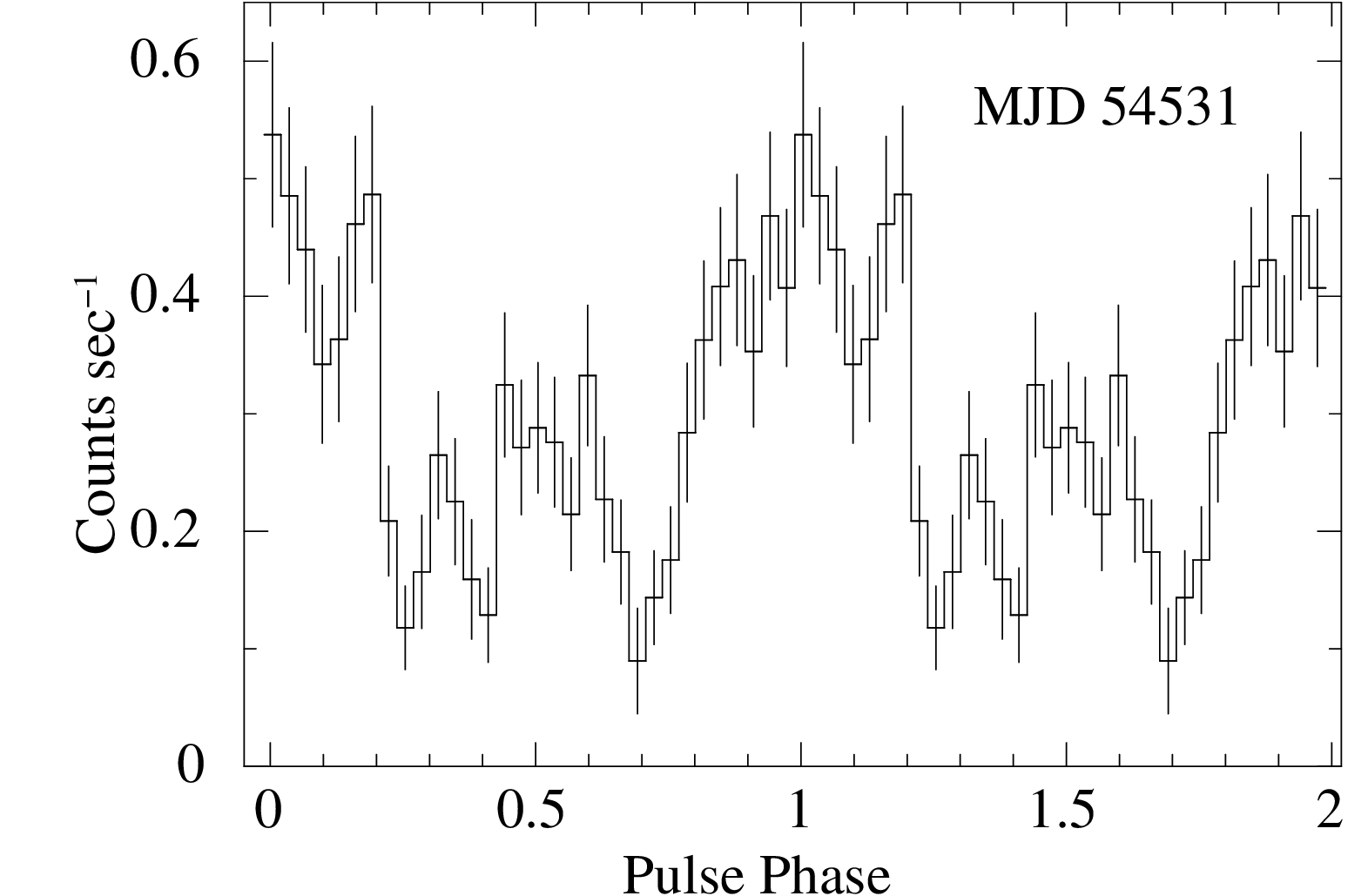}
\includegraphics[scale=0.22, angle=-90]{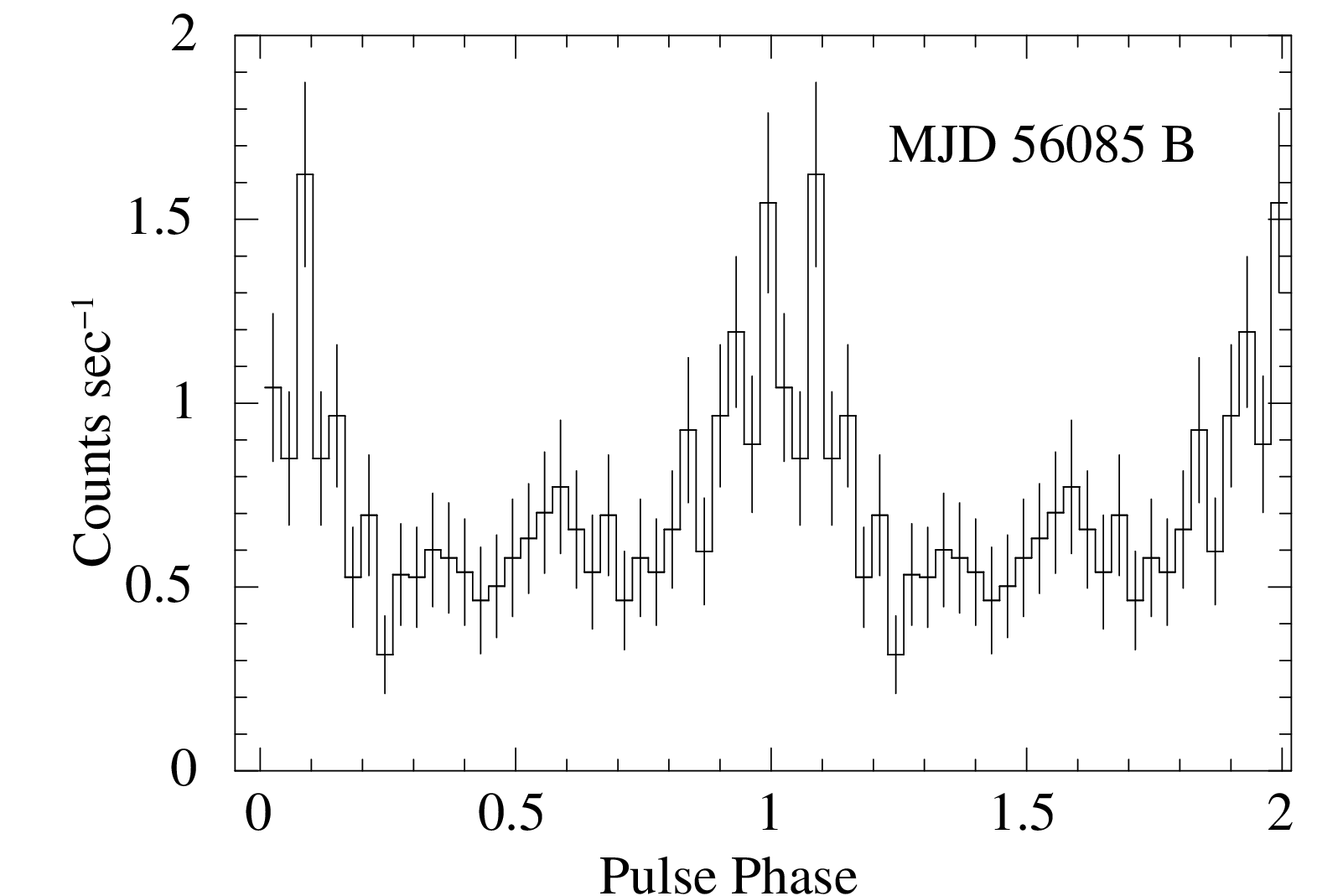}
\includegraphics[scale=0.22, angle=-90]{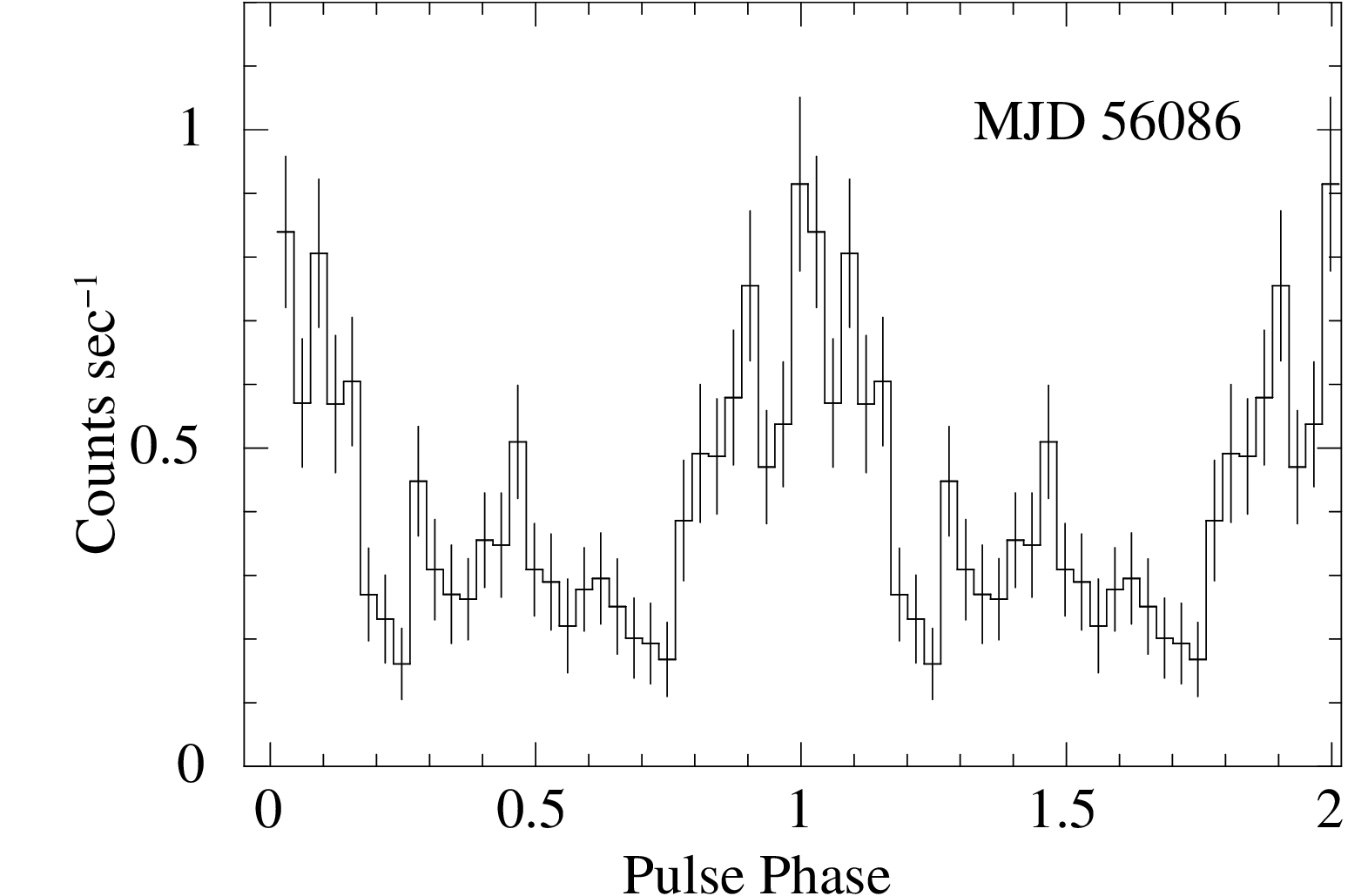}
\includegraphics[scale=0.22, angle=-90]{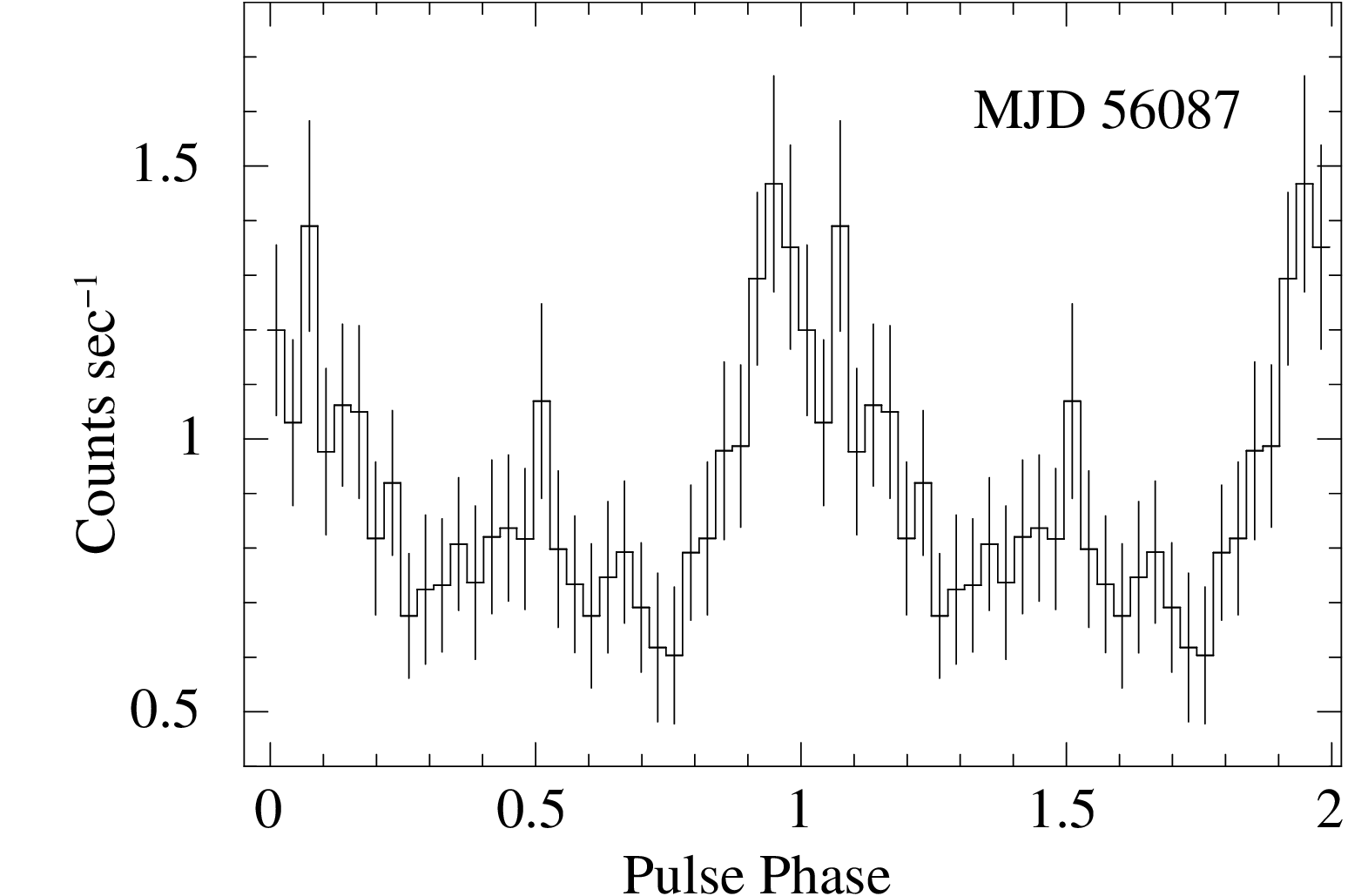}
\includegraphics[scale=0.22, angle=-90]{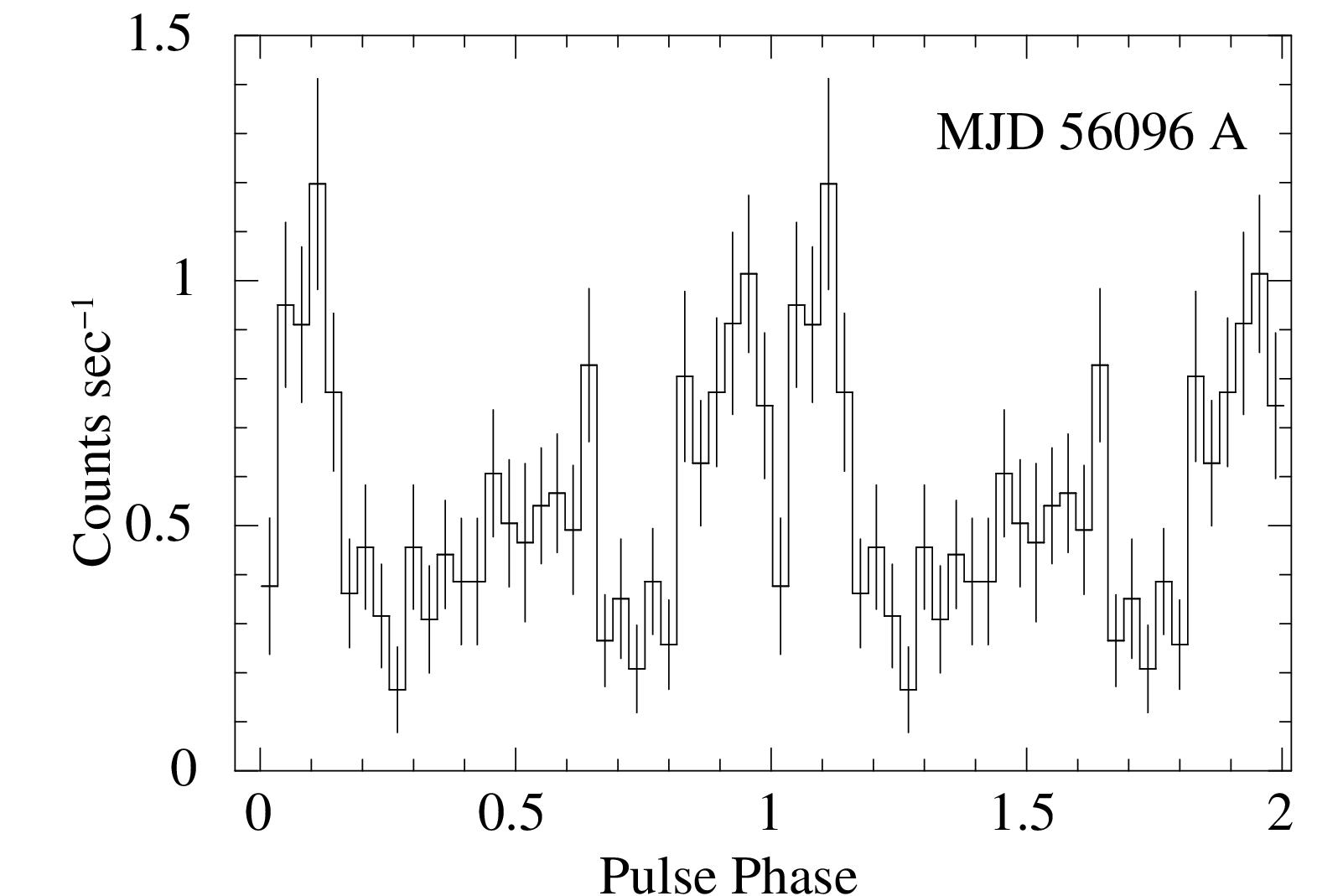}
\includegraphics[scale=0.22, angle=-90]{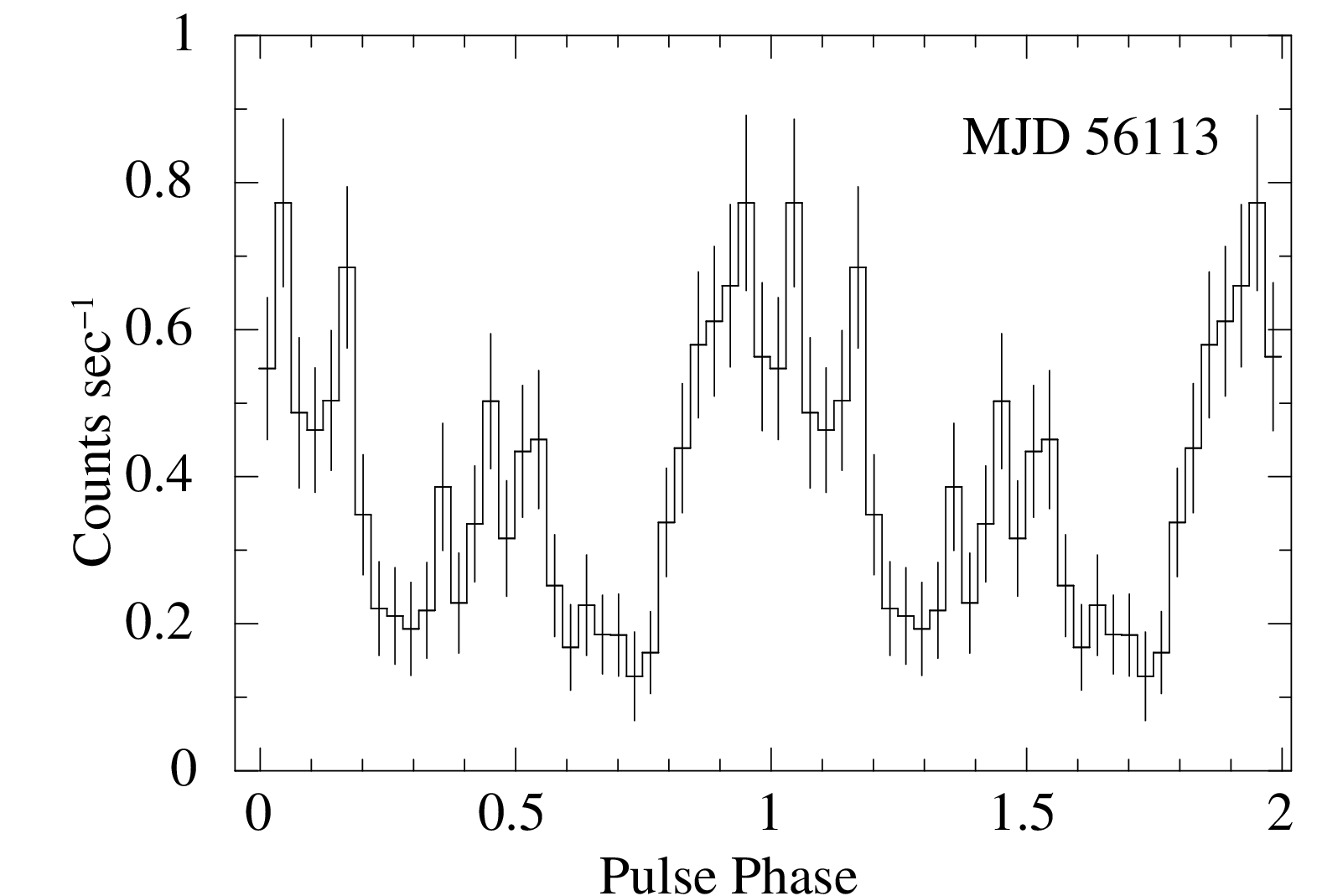}
\includegraphics[scale=0.22, angle=-90]{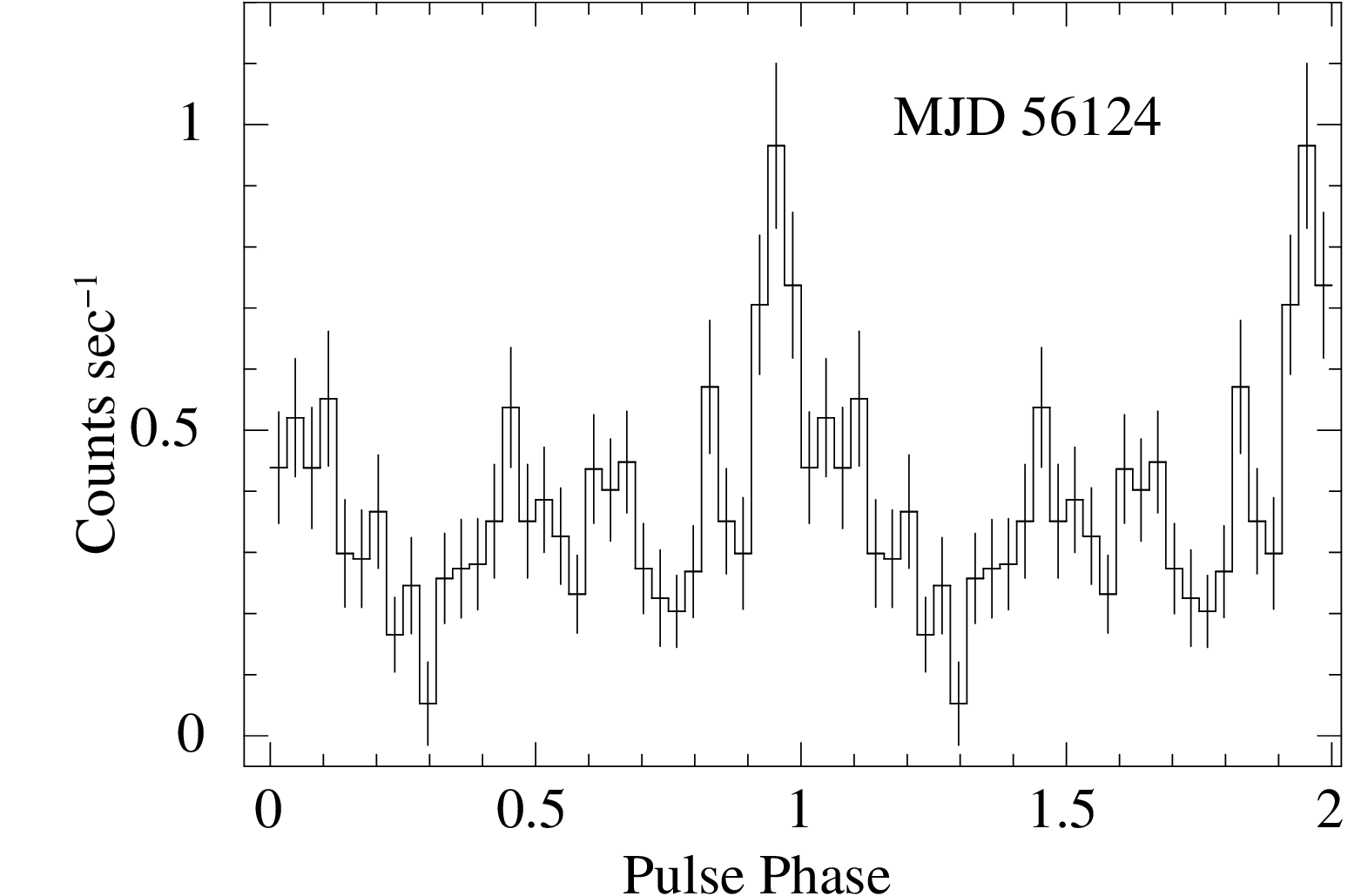}
\includegraphics[scale=0.22, angle=-90]{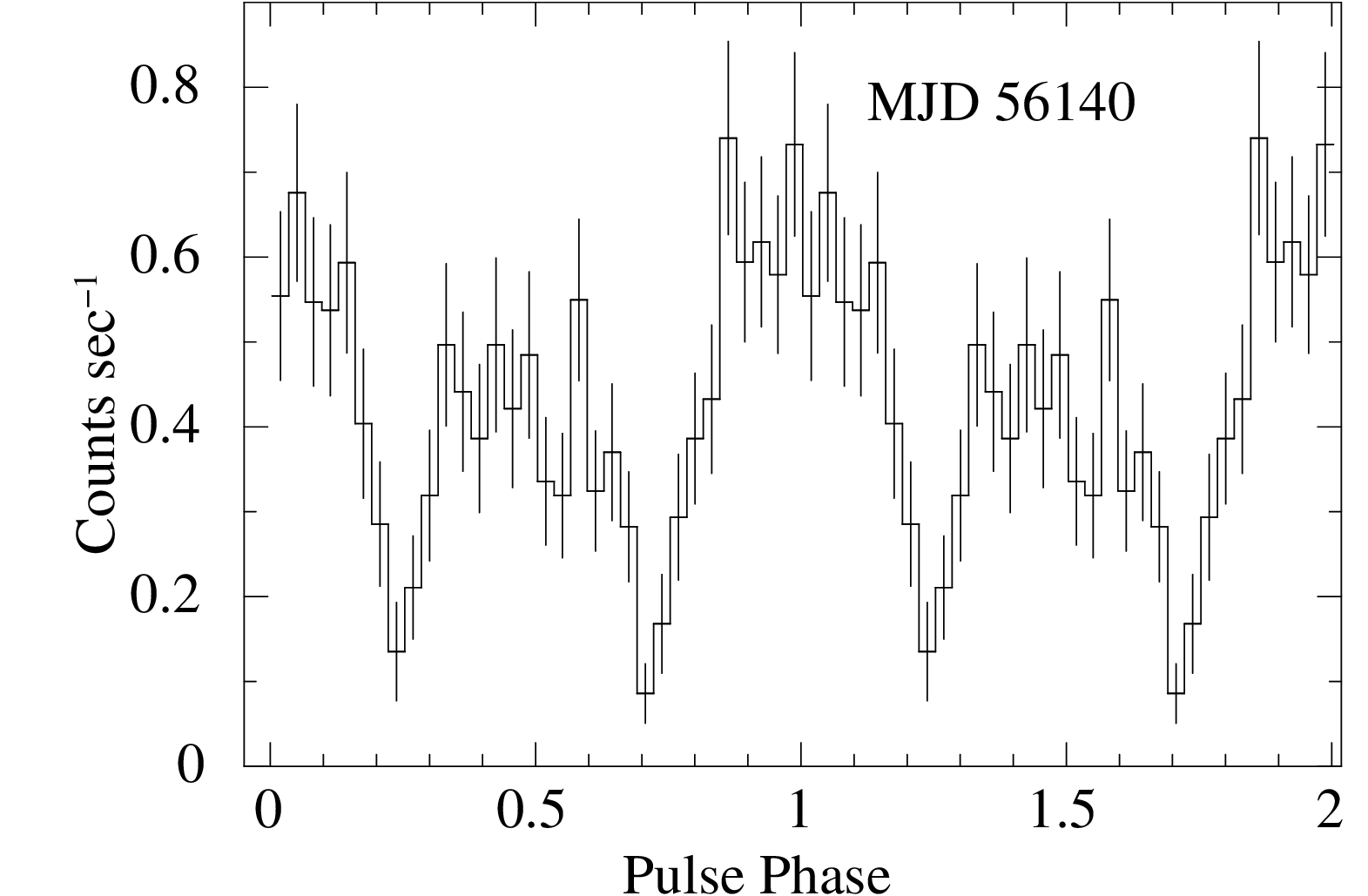}
\includegraphics[scale=0.22, angle=-90]{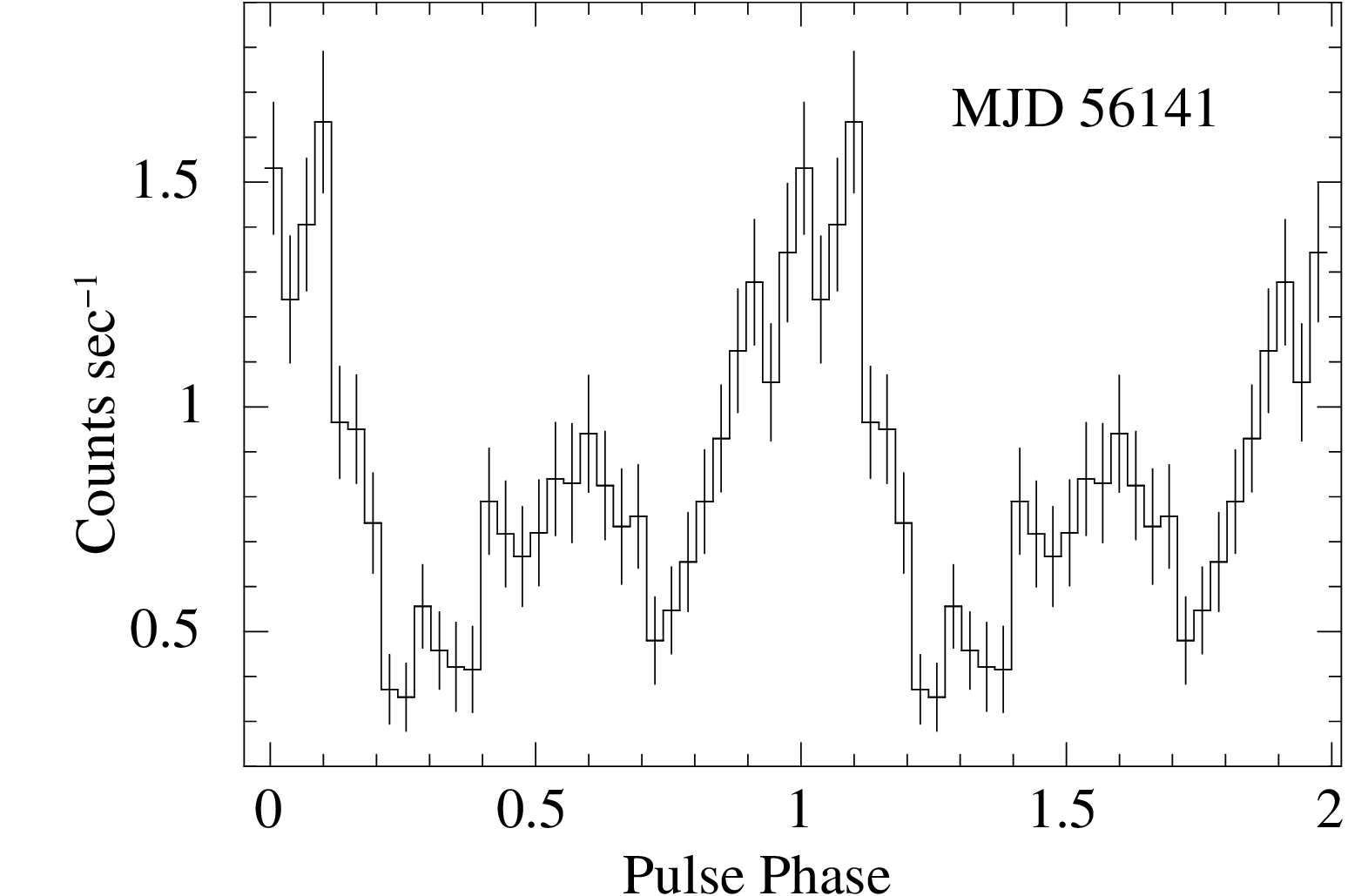}
\includegraphics[scale=0.22, angle=-90]{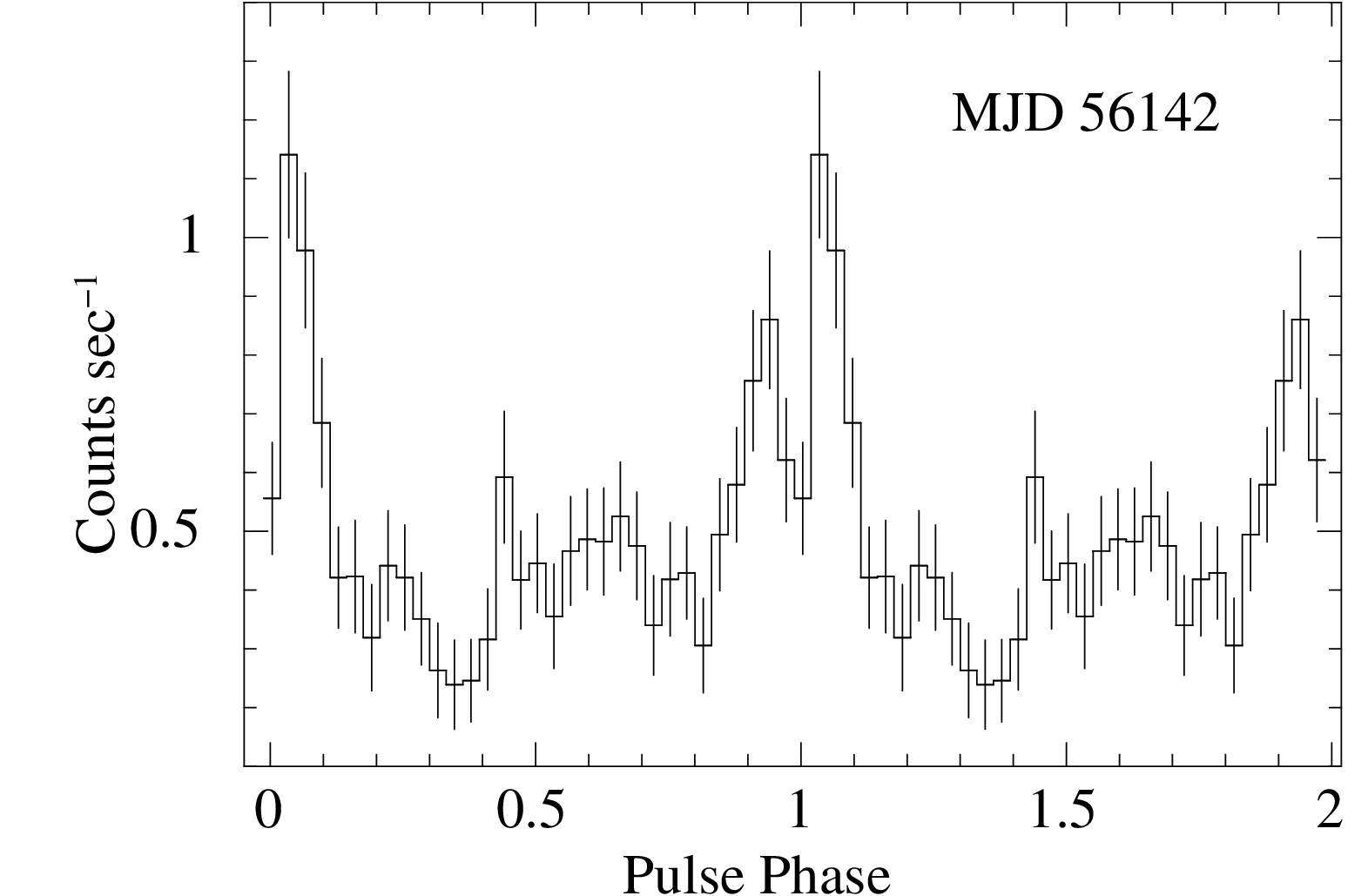}
\includegraphics[scale=0.22, angle=-90]{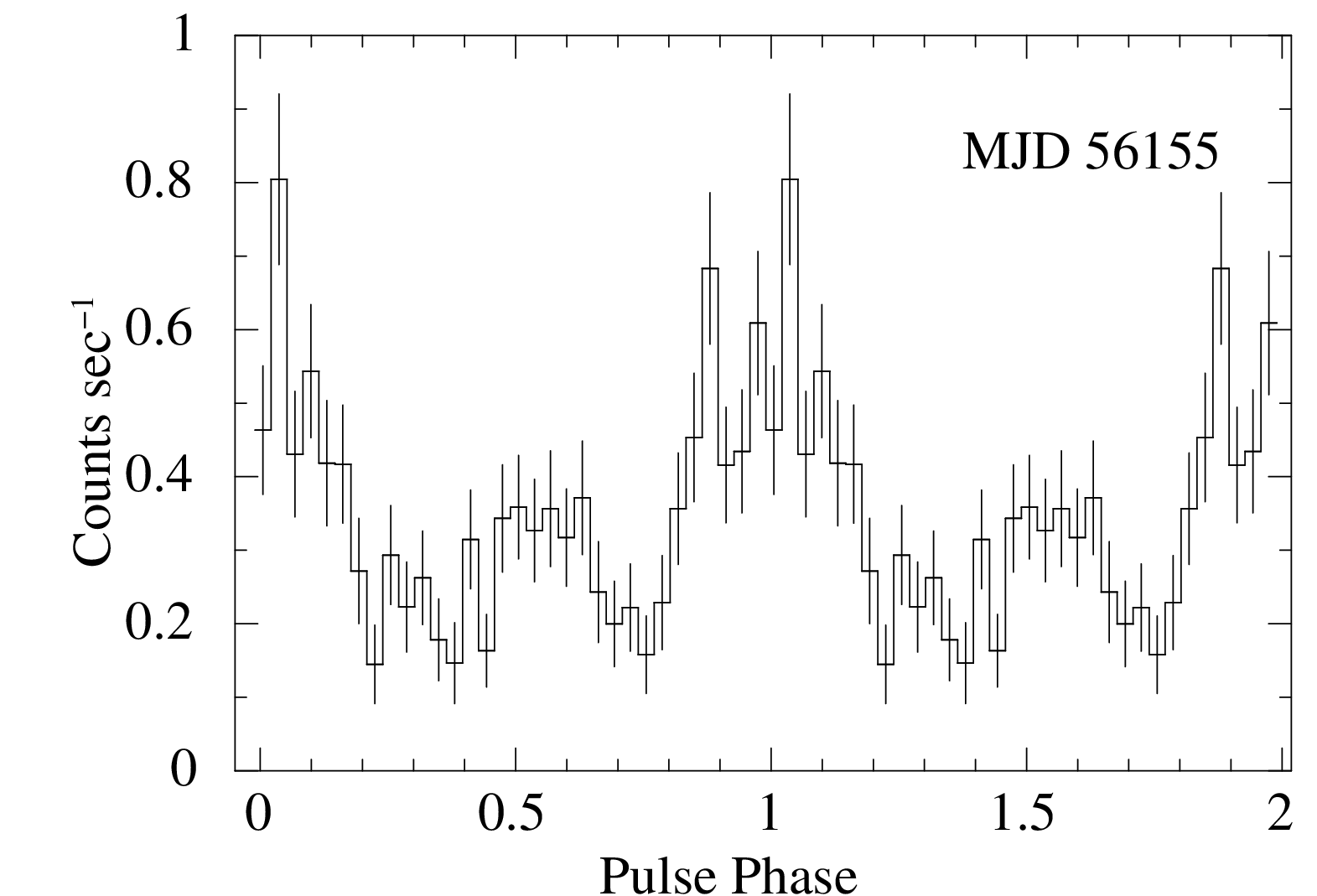}
\includegraphics[scale=0.22, angle=-90]{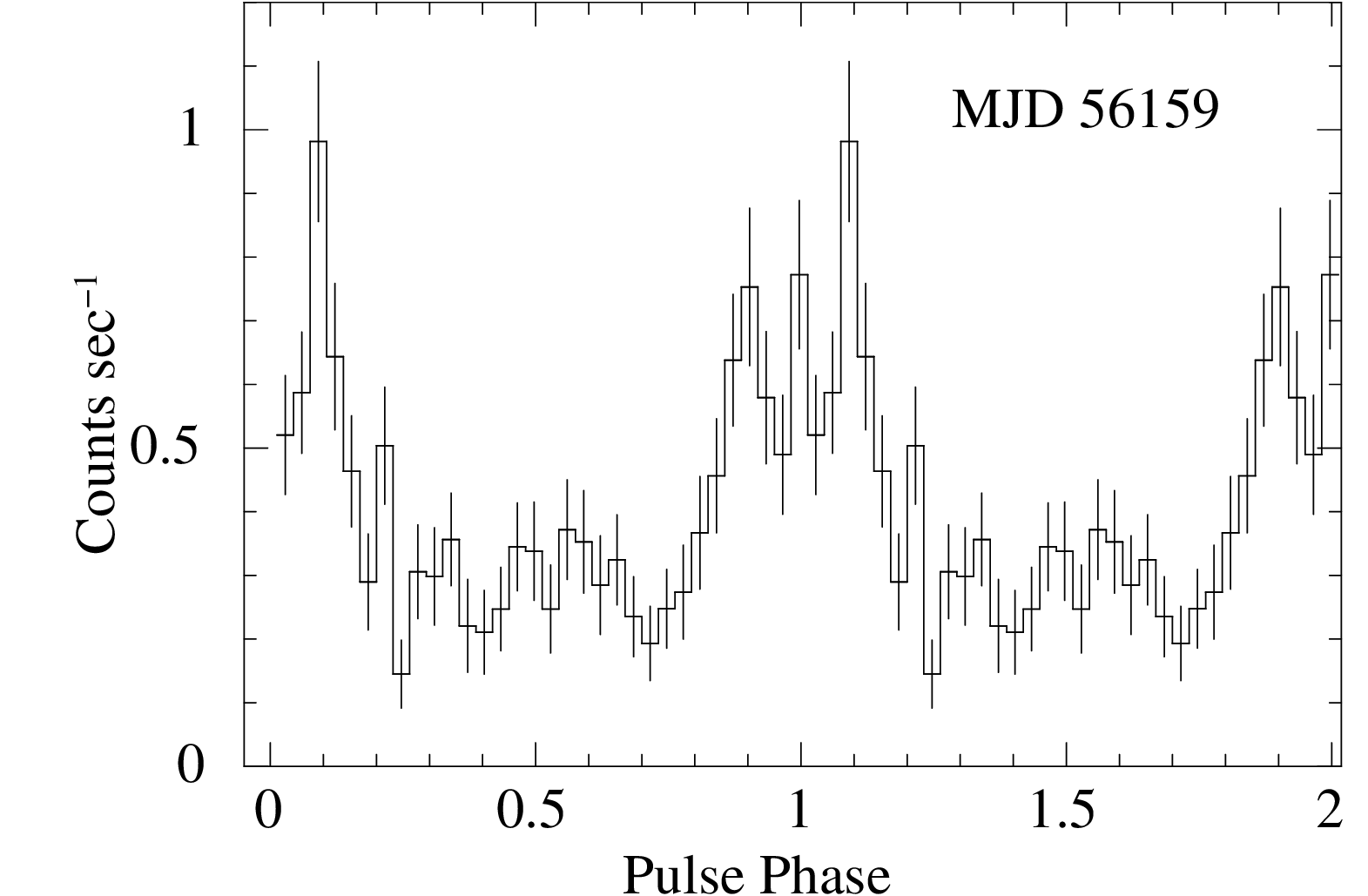}
\includegraphics[scale=0.22, angle=-90]{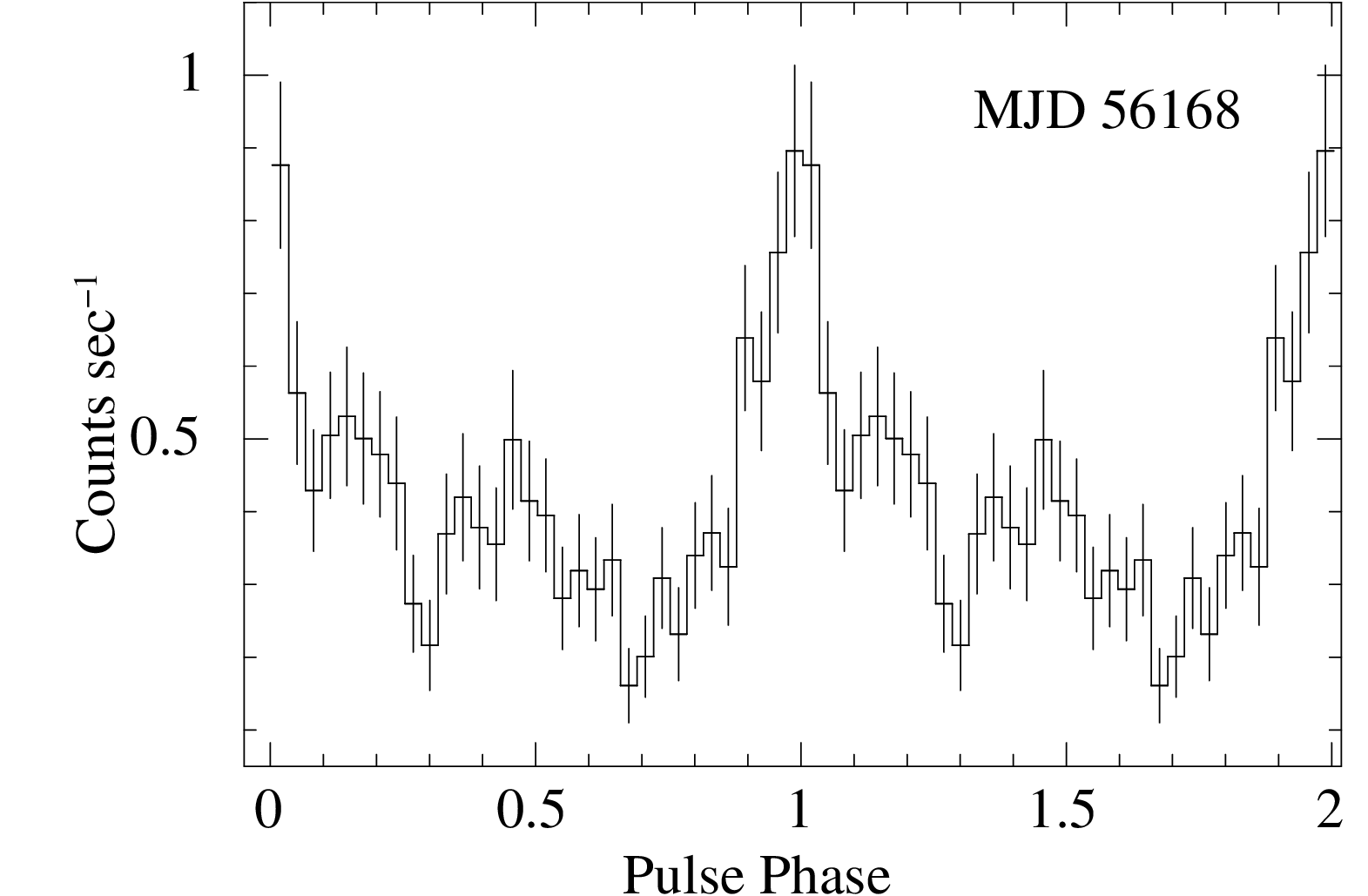}
\includegraphics[scale=0.22, angle=-90]{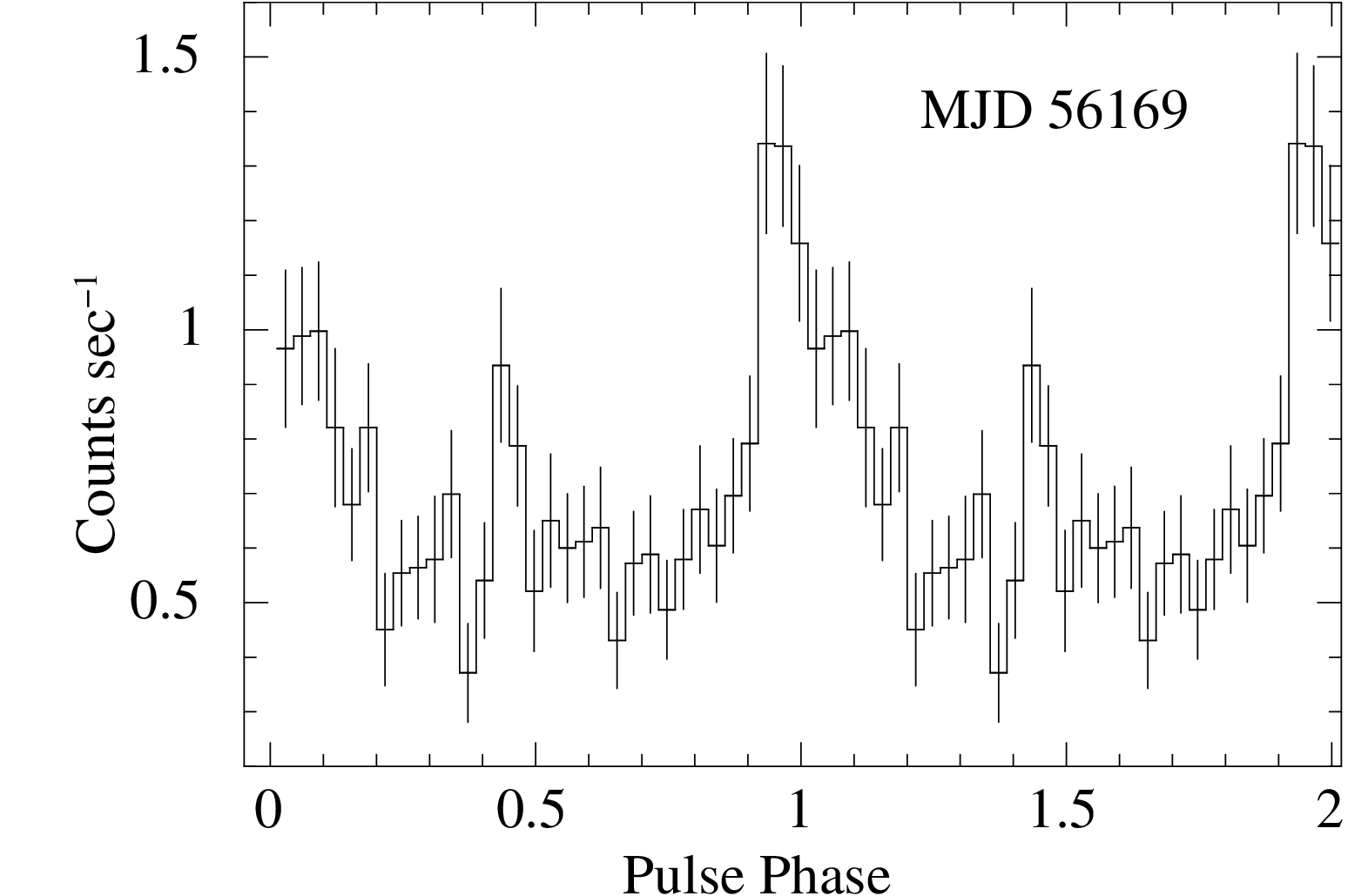}

\caption{Pulse profiles of 15 observations with clear detection of pulsation, folded with their estimated pulse periods and with 
32 phasebins/period. Main peaks of all the profiles have been aligned at phase 1.0. The MJD of each observations are labelled inside each panel.}
\label{fig1}
\end{figure*}

\begin{figure*}
\centering
\includegraphics[scale=0.45, angle=-90]{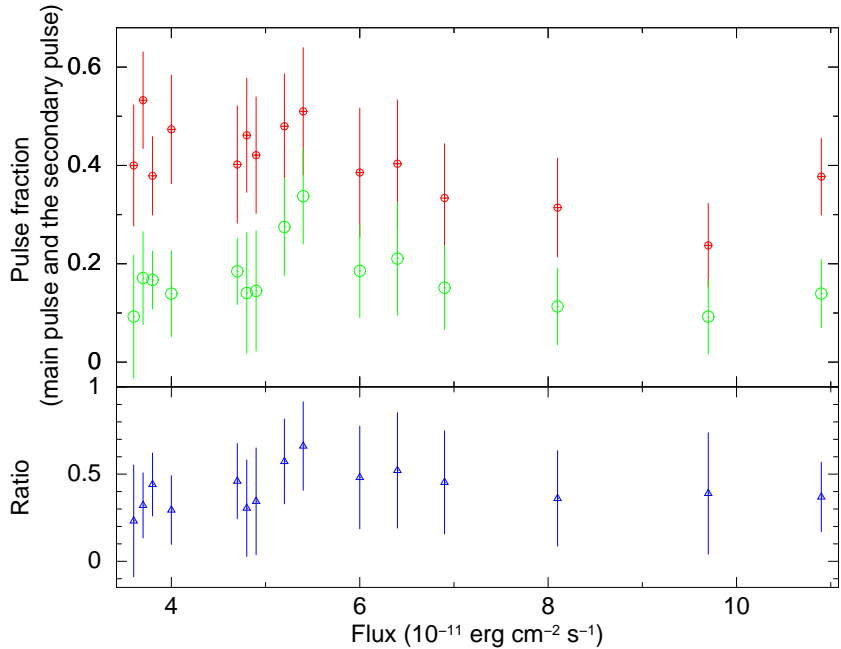} 
\caption{Top panel: Pulse fraction of the main or primary (red) and secondary pulse (green) plotted as function of flux. Bottom panel:
Ratio of the pulse fraction of the primary to the secondary  pulse fraction plotted as function of flux.}
\label{pulse_fraction}
\end{figure*}

\subsection{Pulsation  Analysis}
We searched for pulsations with the {\small FTOOL} task {\small EFSEARCH} for all the observations in which the source 
is clearly visible in the image and the total number of source photons were more than 600. {\small EFSEARCH} results of all 
these lightcurves gave a maximum $\chi^{2}$ of greater than 100 for 32 phasebins per period indicating a clear detection of 
the X-ray pulses. Table 1 shows the exposure time, 
total number of photons in 60$''$ source region, average count rate, pulse period and orbital phase of all observations, along with the
observations which were divided into two parts mentioned in Section 2, arranged in the ascending order of the total number of source photons.
\par
Light curve  of each observations with pulsation detected were folded with corresponding pulse periods. Folded pulse profile of the 
15 observations with pulsation detected are shown in Figure \ref{fig1}. The pulse profiles have been aligned such that the phase of the main
peak of each profile lies at 1.0.
As seen in Figure \ref{fig1}, most of the pulse profiles 
show a double peaked structure, with a possible indication of variation in relative intensity of the peaks. Only for the purpose of 
comparing  the strengths of the two peaks, we  fitted each of these pulse profiles with two gaussian,
one for the primary pulse and the other for the secondary along with an unpulsed component. 
We define pulse fraction of the two peaks as 
the fractional area of the two gaussians. We obtained pulse fraction for both the peaks and plotted them in the top
panel of Figure \ref{pulse_fraction} along with their ratio i.e. the relative pulse fraction of the secondary to the primary in the 
bottom panel as function of flux. We see that while the overall pulse fraction has a weak negative correlation with the flux, the relative pulse 
fraction of the two peaks is nearly constant. 

\begin{table*}
 \label{table}
\caption{ Log of observations with exposure time,  total number of source photons, average count-rate, pulse period 
and orbital phase. A and B refers to the observations which were split as mentioned in Section 2.}
  \centering
 \begin{tabular}{|c|c|c|c|c|c|c|}
 \hline
 &&&&&&\\

Observation  &Observation &Exposure &Total no    & Average     &Pulse     &Orbital\\
 MJD         &ID 	  &time     & of photons &count-rate   &Period    &Phase\\
             &            &         & in source  &             &          &\\
             &            & (sec)   & region     &(counts/sec) & (sec)    &\\
          
 \hline
  56143 &00035720023&	929& 	6&	0.01		 	  &-&0.02 - 0.03\\
  56098 &00035720011&	874& 	8&	0.01		 	  &-&0.06 - 0.08\\
   56085 A &00035720005 A&442	&	16 &	0.04 		  & - & 0.13\\
56171 &00035720034&	2025 & 	22& 	0.01			  &-&0.92 - 0.94\\
56144 &00035720024&	2093 &	26 & 0.01			  &-&0.02 - 0.03\\
 56125 &00035720017& 	2035&	29 & 0.01		 	  &-&0.86 - 0.89\\
 56089 &00035720009&	989 &	42& 0.04		&-&0.25\\
 56157 &00035720027&	1913&	50 & 0.03		&-&0.03 - 0.05\\
 56156 &00035720026&	2035&	81 & 0.04		&-&0.73 - 0.75\\ 
 56167&00035720030&	1975 &	99 &	0.05		&-&0.11 - 0.13\\
 56170 &00035720033&	1988 &	126&	0.06	      &-&0.69 - 0.70\\
  55751        &00035720004 &692   & 	127   & 0.18     & -&0.16\\
  54141 B &00035720001 B&3798	 &142	 &	0.03	               &-&0.83 - 0.91 \\
   56128 &00035720019&	1231& 	180&	0.15		&-&0.48 - 0.53\\
  56088 &00035720008&	2020 & 	204 &	0.10		&-&0.73 - 0.75\\
  56118 &00035720015&	1878 &	218&  0.12		&-&0.48 - 0.5\\
  56126 &00035720018&	1898 &	252 & 0.13		&-&0.20 - 0.22\\
  56100 &00035720013&	2170 &	282 & 0.13		&-&0.36 - 0.42\\
  56158 &00035720028&	1523 &	376 &	0.25		&-&0.16\\
   56096 B &00035720010 B&327	&	415&	1.18		   & - &0.63\\
   56099 &00035720012&	1121& 	417&	0.39		&-&0.17 - 0.23\\
    56096 A &00035720010 A&1059	&	620 &	0.56		        &140.12$\pm{0.01}$ &0.58 - 0.61 \\
   56085 B &00035720005 B&852	& 	699  &	0.73	        	      &141.31$\pm{0.01}$ & 0.28\\
   56113 &00035720014&	1920&	719 & 0.41			  &139.69$\pm{0.01}$ &0.30 - 0.31\\
   56155 &00035720025&	2143 &	772 & 0.36		 	  &140.01$\pm{0.01}$&0.44 - 0.45\\
56124 &00035720016&	1855 &	780 & 0.42		    &143.35$\pm{0.01}$&0.69 - 0.72\\
   56086 &00035720006&	1873 &	796 &	0.43		    &139.75$\pm{0.01}$&0.30 - 0.31\\
   56159 &00035720029&	1968 &	822&	0.42			  &140.01$\pm{0.01}$&0.36 - 0.39\\
   56140 &00035720020&	1955 &	876 & 0.45		    &139.66$\pm{0.01}$&0.16 - 0.17\\
   56168 &00035720031&	2038 &	900& 0.44		    &140.14$\pm{0.01}$&0.23 - 0.25\\
   54531 &00035720002&	2914 &	909 &	0.31		    &139.66$\pm{0.01}$&0.17 - 0.23\\
   56142 &00035720022&	1960 &	1035 & 0.53		    &139.48$\pm{0.01}$&0.69 - 0.72\\
   56087 &00035720007&	1446 &	1329&	0.92		    &143.25$\pm{0.01}$&0.53 - 0.55\\ 
   56169 &00035720032&	1865 &	1370 &0.74		    & 140.20$\pm{0.01}$&0.47 - 0.48\\
     54141 A &00035720001 A&5789	 &	1617 &0.27	    &139.95$\pm{0.01}$&0.67 - 0.77 \\
   56141 &00035720021&	1970 & 	1729& 	0.88		    &139.87$\pm{0.01}$&0.48 - 0.52\\
   &&&&&&\\
 \hline
 \end{tabular}
 \end{table*}

\begin{figure*}
\centering
  \includegraphics[scale=0.4,angle=-90]{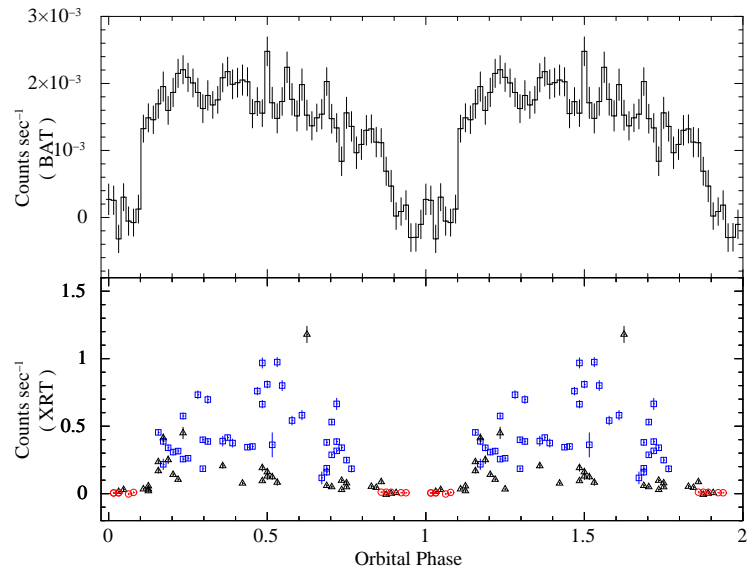}

 \caption{{\it Top panel:} Orbital intensity profile of IGR J18027--2016 obtained by folding {\it Swift}--BAT light-curve with 
 orbital period of 394843 sec. {\it Bottom panel:} {\it Swift}--XRT lightcurves modulo same orbital period 
 in three colours: pulsation detected   where number of source photons is greater than 600 -- blue, number of source photons is 
 less than 600  -- black; faint -- red.}
 \label{orbit}
\end{figure*}

\begin{figure*}
\centering
\includegraphics[scale=0.25, angle=-90]{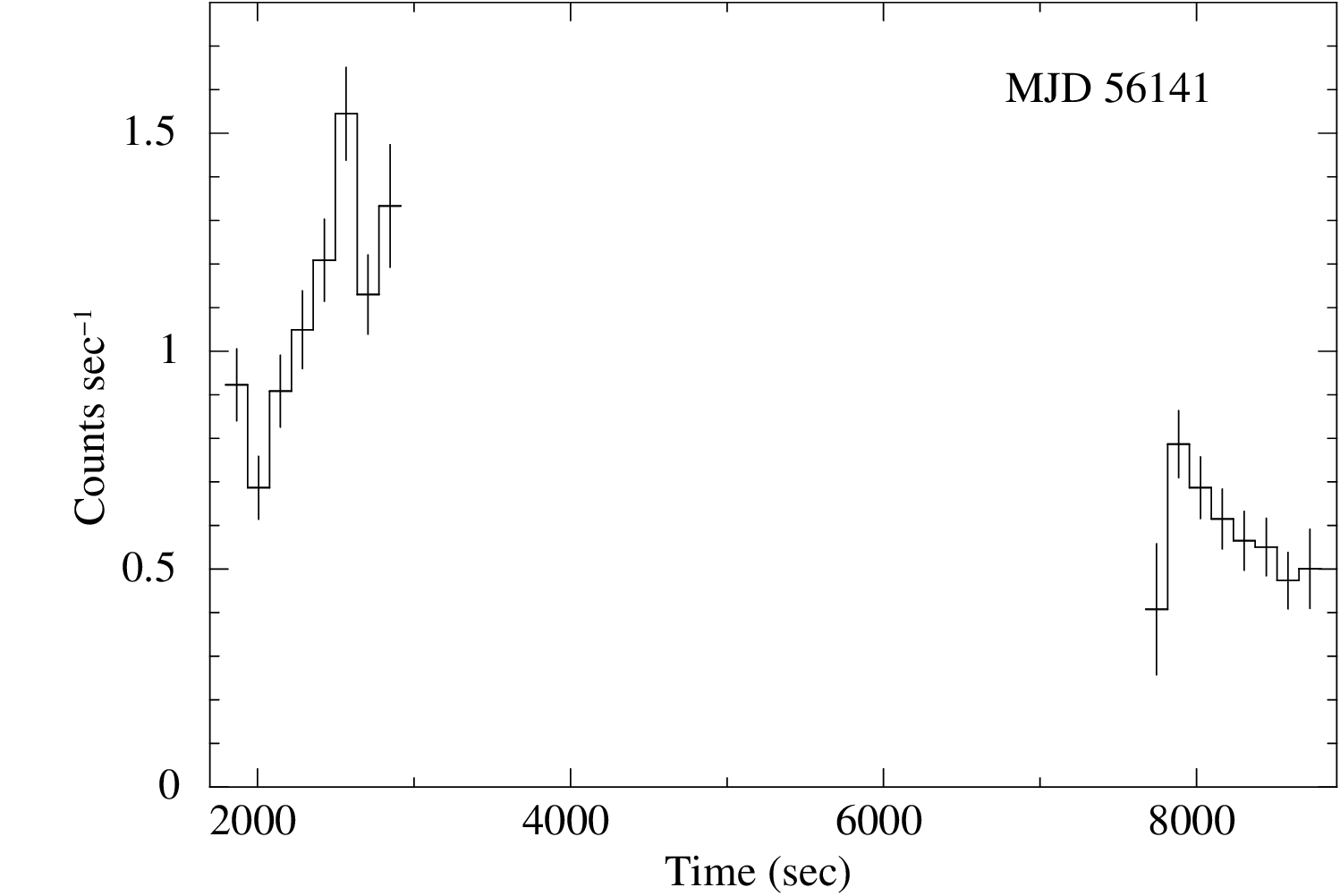}
\includegraphics[scale=0.25, angle=-90]{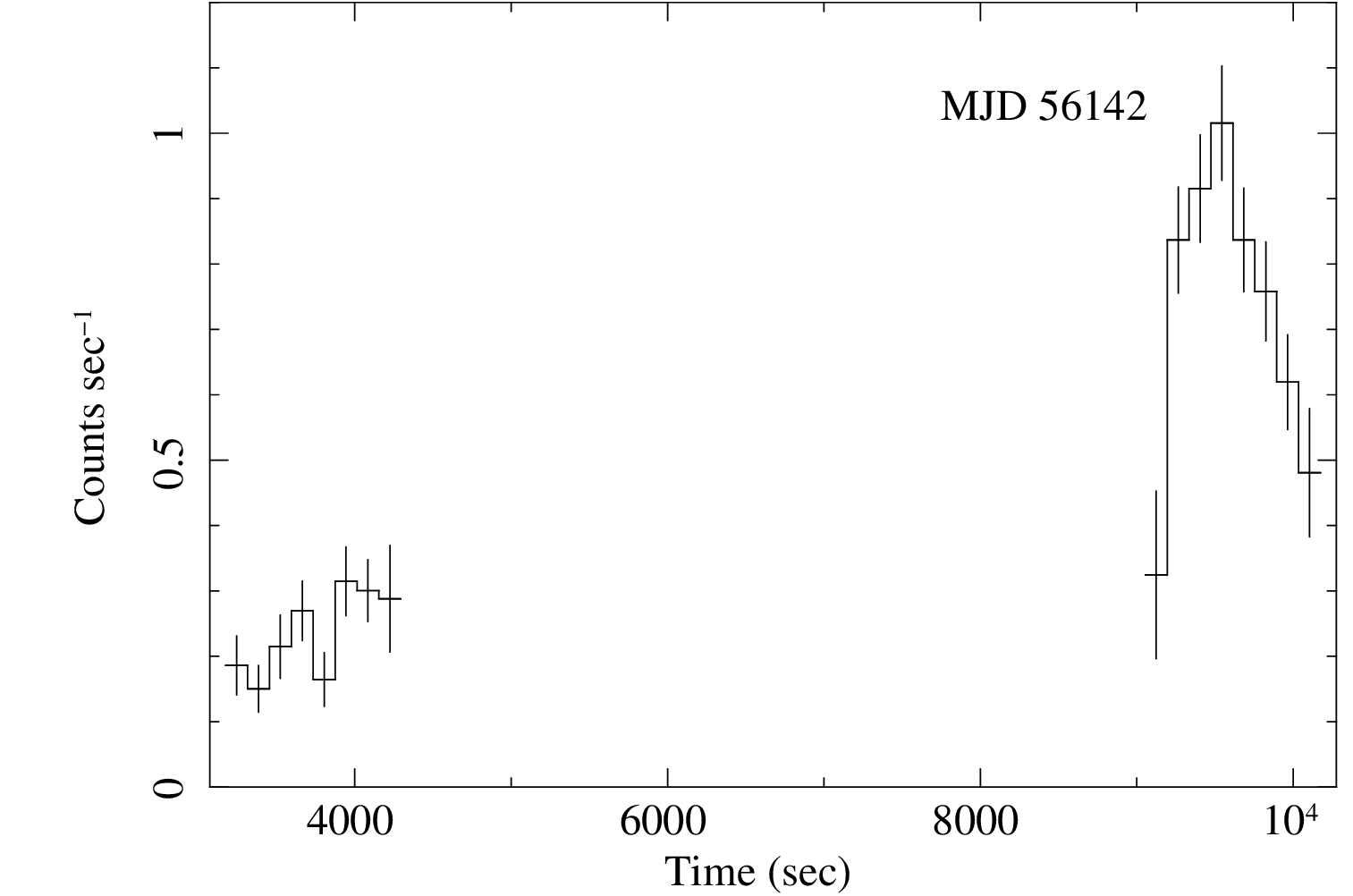}
\includegraphics[scale=0.25, angle=-90]{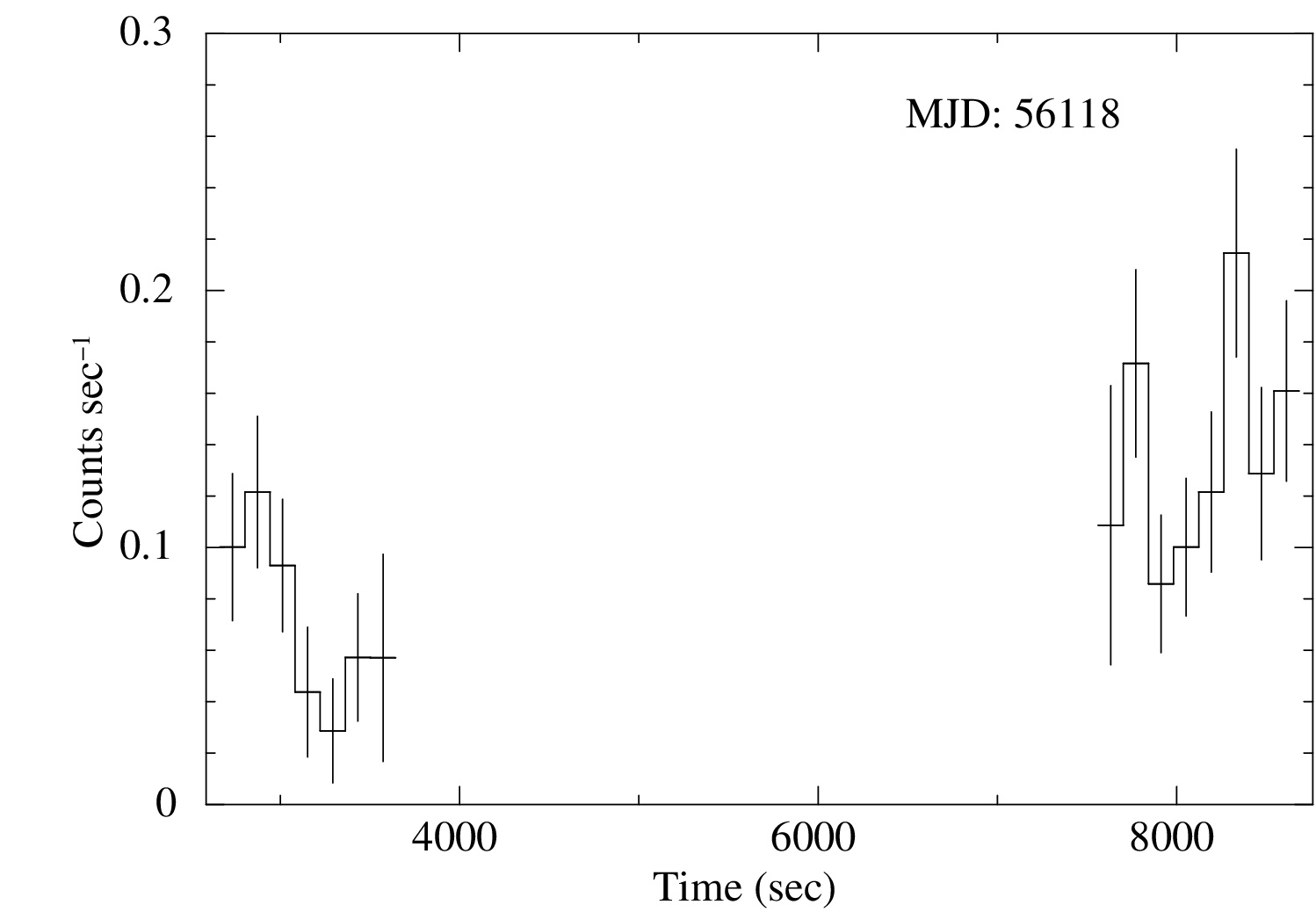}
\includegraphics[scale=0.25, angle=-90]{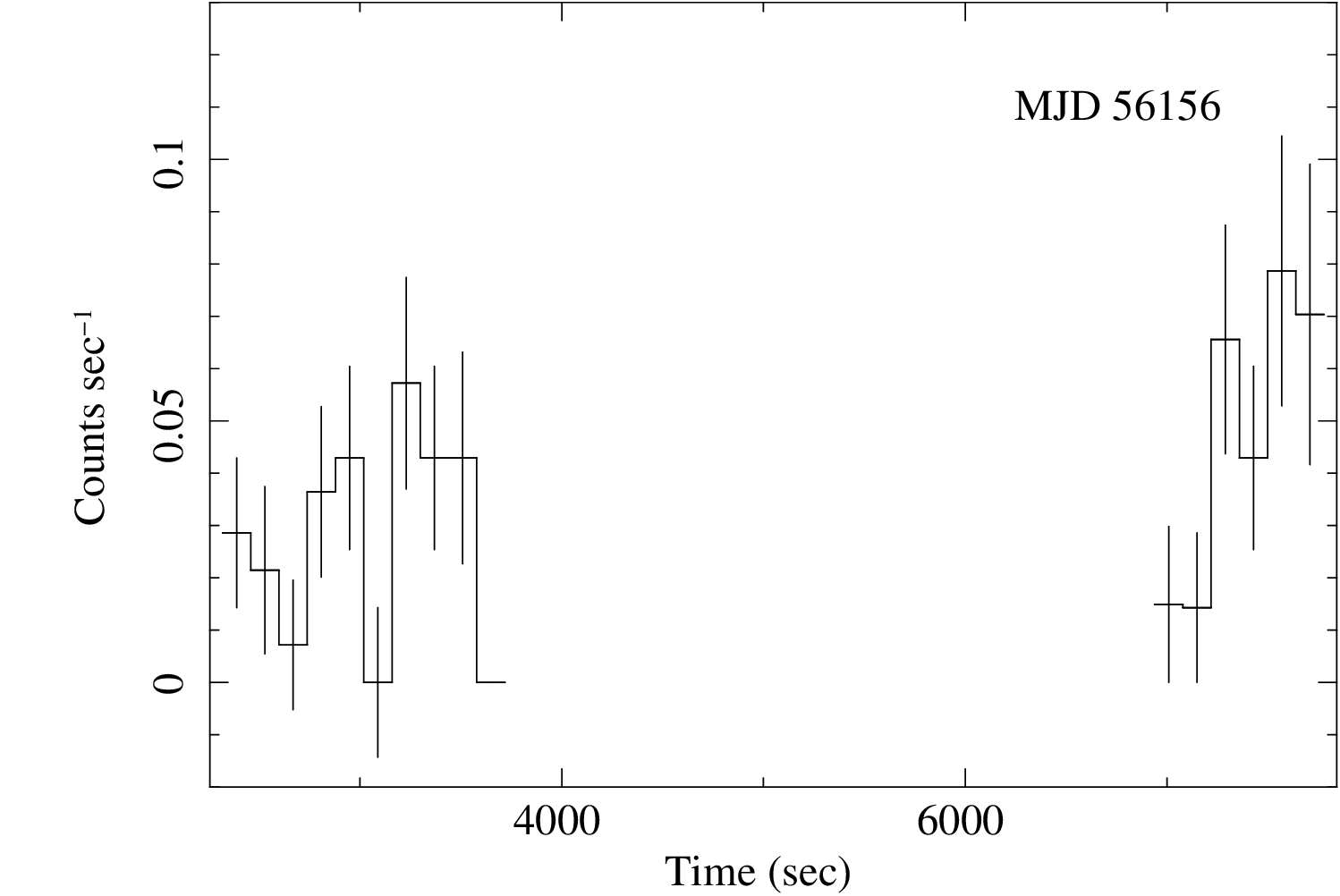}
\includegraphics[scale=0.25, angle=-90]{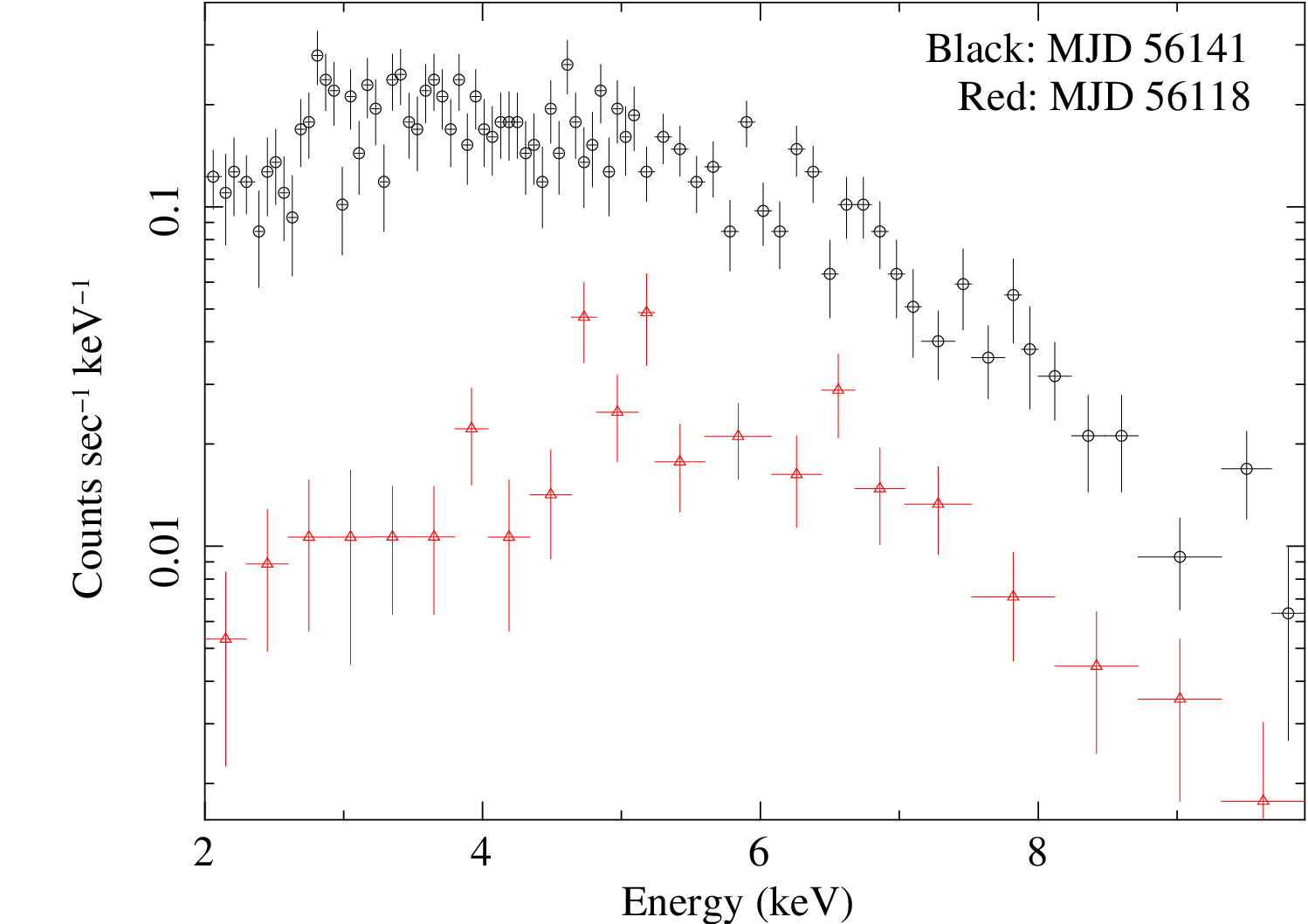}
\includegraphics[scale=0.25, angle=-90]{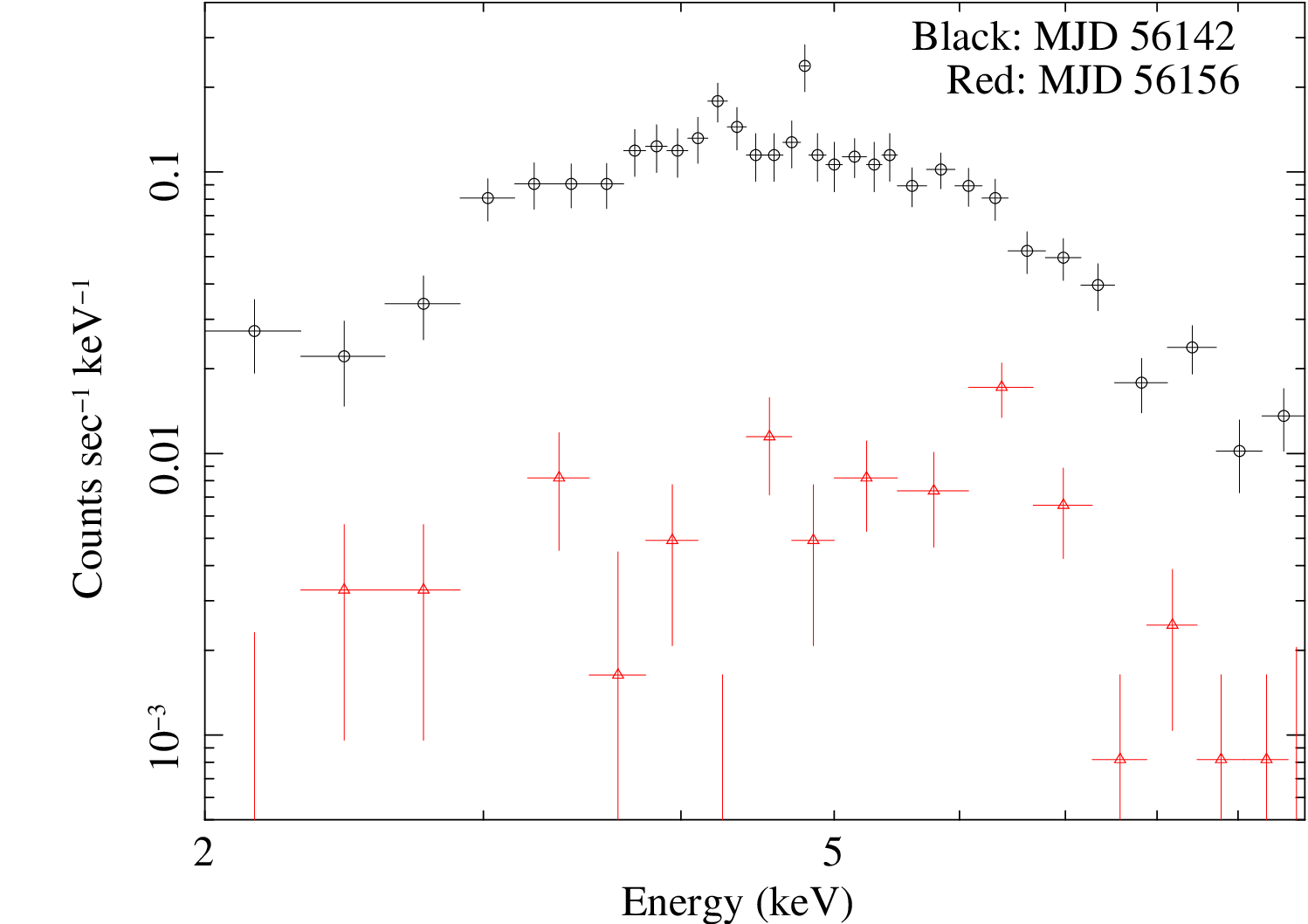}
\caption{{\it Left panel}: The top and middle panel show the light-curves of two observations centered at orbital phase 0.48-0.52. 
In the top panel, the light-curve of an observation (MJD 56141) shows a high count-rate whereas the middle panel 
shows another observation (MJD 56118) in the same orbital phase range, having a low count-rate. The lower panel is the plot of 
spectra of these two observations which bring out the fact that in the same orbital phase range, the X-ray intensities vary by a 
factor of $\sim$10. {\it Right panel}: Same is shown for another set of observations (MJD 56142 and MJD 56156) centered at orbital 
phase range 0.69-0.75, but showing a large change in X-ray intensities.}
\label{split}
\end{figure*}

\subsection{Orbital period analysis}
 We have searched for orbital period in the {\it Swift}--BAT light curve with the {\small FTOOL} task {\small EFSEARCH} and found it 
 to be 394843 sec (4.57 days; similar to P$_{orb}$ determined by \citealt{hill2005,jain2009}).
  We then folded the {\it Swift}--BAT lightcurve with this orbital period, and in the folded profile (shown in
  the top panel of Figure \ref{orbit}), we can see an eclipse for duration of  nearly $\frac{1}{4}$th of the orbital period.
  In the bottom panel of Figure \ref{orbit},   we have plotted the orbital phase-wise X-ray count-rates obtained from all
  {\it Swift}--XRT lightcurves in three colours: pulsation detected   where source photon is greater then 600 -- blue; 
  source photon less than 600  -- black;   faint, i.e. where source could not be seen clearly -- red.  
  To investigate any intensity variations other than the long time   averaged orbital intensity modulation, multiple
  observations with {\it Swift}--XRT during the same orbital phase range are   not averaged here, unlike the orbital 
  profile  shown in Figure 3 in \citet{bozzo2015}.
 \par
 In Figure \ref{orbit}, the bottom panel shows the variability in count-rate of the source in the orbital intensity profile 
 with the pointed {\it Swift}--XRT observations, whereas the {\it Swift}--BAT orbital intensity profile is averaged over
  many orbital cycles, indicating a sub-orbital variability similar to that seen in IGR J16393-4643 \citep{nazma2015} and 
  OAO 1657--415 (\citealt{pp2014,barn2008}). Around orbital phase 0.5, there are multiple {\it Swift}--XRT observations showing significant 
  difference in the count rates. We have shown spectra and lightcurve for two parts of observations carried out in same orbital 
  phase ranges (0.48-0.52, 0.69-0.75) in Figure \ref{split}. The light curve is binned with 140 seconds  
  (close to the spin period of the pulsar) to avoid seeing any effect of the pulse profile related variation in the
  light curve. In the top panel in Figure \ref{split} there are about 140 photons per bin and the variability in the light
  curves is clearly larger than the photon noise (represented by the 1$\sigma$ error bars in each bin). The number of counts 
  per bin in the two light curves shown in the middle panel is smaller and have correspondingly larger uncertainties. 
  No intensity variation can be ascertained in the light curves shown in the middle panles. Spectra of these two observations
  are shown in the lower panel of Figure \ref{split} which bring out the fact that in the same orbital phase range, 
  the X-ray intensities vary by a factor of $\sim$10.

\begin{figure*}
\centering
\includegraphics[scale=0.3, angle=-90]{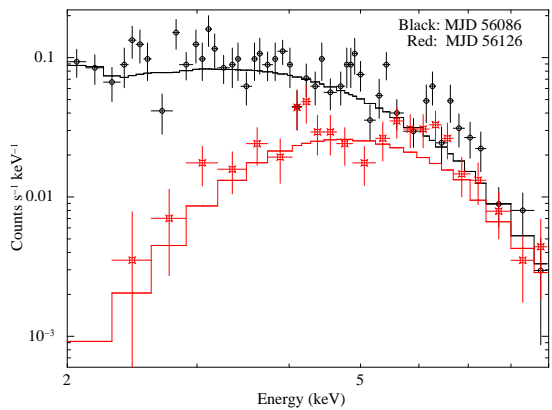}
\caption{Spectra at MJD 56086 AND 56126 are plotted together to show the variation in 
low energy absorption}
\label{two_spec}
\end{figure*}

\subsection{Spectral Analysis}
\begin{figure*}
 \centering
\includegraphics[scale=0.4, angle=-90]{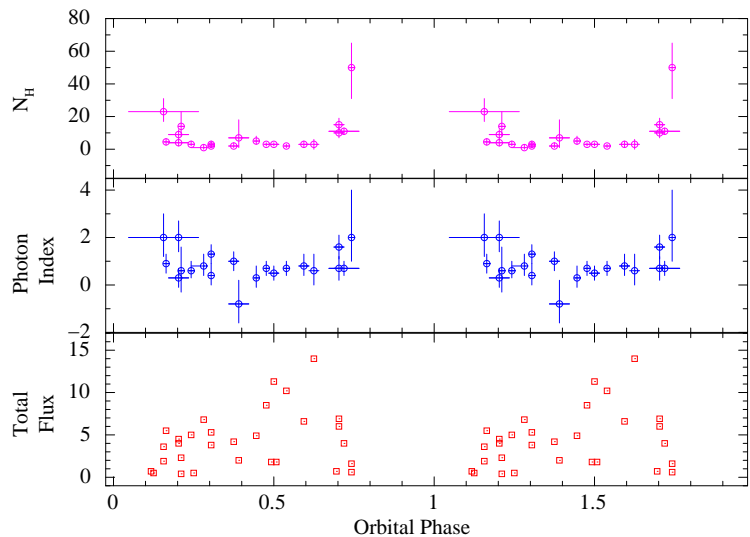}
\caption{Variation of column density of hydrogen (N$_{H}$ in units of 10$^{22}$cm$^{-2}$), photon index ($\Gamma$), 
total flux (F in the units of 10$^{-11}$ ergs.cm$^{-2}$sec$^{-1}$). }
 \label{parm}
 \end{figure*}

We fitted X-ray spectrum for 21 observations, having moderate statistics, using  {\small XSPEC} v12.8.2 in the energy range 2.0--9.0 keV.
Because of limited statistics. The spectra were modelled with a
power law modified by a photoelectric absorption by column density of absorbing matter along our line of sight. 
We have also fitted the spectra from the remaining observations only for the purpose of estimating the total flux.
We have found the equivalent column density of hydrogen (N$_{H}$) in the range of
10$^{22}$--10$^{23}$ cm$^{-2}$. 
We have plotted two spectra in Figure \ref{two_spec} obtained at MJD 56086 and 56126 to show the variation in the absoption at low energies. 
The spectra clearly indicate large difference in column density. 
Flux during the out-of-eclipse orbital phase are found to be in the range of (0.4--14) $\times$10$^{-11}$ ergs cm$^{-2}$ sec$^{-1}$.
We have plotted the spectral parameters N$_{H}$, $\Gamma$ and total flux (2.0--9.0 keV) from the system in Figure \ref{parm}. 

\section{Discussion}
In this work, we have analyzed all available {\it Swift}--XRT observations of the HMXB source IGR J18027--2016, to study its pulsations 
and variability characteristics. Pulsations have been detected in all the observations having a total number of source 
photons greater than 600 (Table 1) and the light-curves of these observations were folded with the pulse periods to 
create pulse profiles (Figure \ref{fig1}). Some of the pulse profiles are found to have double peaked structure. We therefore 
carried out an analysis of the pulse profile and determined the pulse fractions of the two peaks. We have plotted these pulse
fractions and their ratio (primary pulse fraction to the secondary) in Figure \ref{pulse_fraction} as function of flux and
find no evidence for significant variation in the pulse profiles over a factor $\sim$3 variation in flux.
\par
 The pulse profiles of accreting X-ray pulsars show strong energy dependence \citep {nagase1989}, usually having simpler pulse
 profile at higher energies ($>$10 keV) and complex profile at low energies, often resulting due to phase locked absorption.
 However in a given energy band, most persistent HMXB pulsars {\it i.e} sources with supergiant companion have pulse profiles
 that are stable over long periods of time (Vela X--1 -- \citealt{krk1999}, \citealt{chand2013}).  
The transient pulsars, on the other hand, show very strong time/luminosity dependence of the pulse profile, 
which can be attributed to the changes in the structure of the X-ray emission region (accretion column) during the transient 
phase \citep {jincy2011}. The limited pulse profile changes in IGR J18027--2016 is consistent with other persistent HMXBs.
\par
The long term averaged orbital intensity profile of this source created with {\it Swift}--BAT light-curves is smoothly varying,
having an eclipse 
lasting for about 1/4 th of the orbit (top panel of Figure \ref{orbit}). The {\it Swift}--XRT and 
{\it INTEGRAL} light-curves, when averaged also give smoothly varying orbital intensity profiles \citep {bozzo2015, hill2005}. 
However, the {\it Swift}--XRT light-curves, when plotted individually for all the observations as a function of orbital phase, 
shows a significant count-rate variation outside the eclipse (bottom panel of Fig \ref{orbit}).
Within the same observation carried out around orbital phase 0.3-0.5 (MJD: 56085), the X-ray count-rates are found to 
vary by a factor of 36. A maximum count-rate variation (a factor of 48) is shown by two non-eclipse observations at MJD 56085 and MJD 56087. 
These short term variation could also be due to hydrodynamical instabilities. \citet{mano2015} has produced the
hard X-ray variation observed with INTEGRAL-ISGRI and RXTE-PCA with hydrodynamical instabilities predicted by simple model without
considering intrinsic clumping or propeller effect. In some cases like in OAO 1657--415, the variations in X-ray intensity may also 
arise due to the accretion onto the compact object by inhomogeneous clumpy winds  \citep{oski2013, pp2014, barn2008}. 
\par
In the present work, we detect several low X-ray intensity episodes 
(For {\it e.g} in orbital phases 0.5 and 0.7) in the supergiant HMXB IGR J18027--2016, indicating these episodes to be either off-states 
like episode similar to Vela X--1 \citep{doro2011} or possibly the presence of clumpy wind like OAO 1657--415 \citep{pp2014, barn2008}. 
From these XRT observations, we cannot distinguish from either of these two or other scenarios. 
\par
X-ray spectra were extracted for 21 {\it Swift}--XRT observations, having moderate statistics to fit with a simple power-law model, modified 
for photo-electric absorption. spectra of other observations with limited statistics were also fitted with the same models 
just for the purpose of estimating total flux. The value of absorption column density N$_{H}$ is as high as  
$5 \times 10^{23}$ cm$^{-2}$, which is similar to the values obtained by \cite{hill2005,walter2006}, and indicate a 
dense circumstellar environment around the neutron star. From Figure \ref{parm}, we see an increase in N$_{H}$ just before and after 
the eclipse, similar to that seen in 4U 1538--52 \citep{roca2015, mukherjee2006}.

\par
IGR J18027--2016 presents an interesting case of a supergiant source showing evidence of low X-ray intensity states,
similar to well known sources like Vela X-1.
Detailed X-ray timing and spectroscopic observations of IGR J18027--2016 at various orbital phases with other X-ray missions 
would be useful to understand the nature of these low intensity states.

\vspace{15mm}
\textbf{ACKNOWLEDGEMENT}\\
We thank the referee Roland Walter for suggestions that helped us to improve the paper.
The data used for this work has been obtained through the High Energy Astrophysics Science Archive (HEASARC) On-
line Service provided by NASA/GSFC. We have also used the public light-curves from {\it Swift}--BAT site.
  
\bibliography{bibtex}{}
\bibliographystyle{mn2e}

\end{document}